\documentclass[12pt]{article}
\usepackage{verbatim,color,amssymb,bbm,epsfig}
\usepackage{subcaption}
\usepackage[usenames,dvipsnames]{xcolor}
\usepackage{fancyhdr}
\usepackage[authoryear, sort]{natbib}
\usepackage{amsmath}
\usepackage{pifont}
\usepackage{adjustbox}
\usepackage{multirow}
\usepackage{rotating}
\usepackage{hyperref}

\renewcommand\footnoterule{\kern-3pt \hrule \textwidth 2in \kern 2.6pt}

\def\boxit#1{\vbox{\hrule\hbox{\vrule\kern6pt \vbox{\kern6pt \textcolor{blue}{#1}\kern6pt}\kern6pt\vrule}\hrule}}

\def\authorfootnote#1{{\let\thefootnote\relax\footnotetext{#1}}}

\def\bLambda{\boldsymbol{\Lambda}}

\def\bgamma{\boldsymbol{\gamma}}
\def\bZ{\boldsymbol{Z}}
\def\bxi{\boldsymbol{\xi}}
\def\bmu{\boldsymbol{\mu}}
\def\bomega{\boldsymbol{\omega}}

\def\bepsilon{\boldsymbol{\epsilon}}

\begin{document}
\thispagestyle{empty}
\baselineskip=28pt
\begin{center}
%{\LARGE{\bf  A functional joint model with baseline functional covariates: application to objective physical activity data}}
%{\LARGE{\bf  Daily sitting profiles, physical function, and mortality: a joint model with baseline functional covariates}}
%{\LARGE{\bf A joint model for longitudinal and survival outcomes with baseline functional covariates, with application to sitting behavior in older women.}}
{\LARGE{\bf  A functional joint model with baseline functional covariates: linking sitting accumulation patterns to physical function and mortality among older women}}

\end{center}

\baselineskip=12pt

\vskip 2mm
\begin{center}

Luo Xiao\\
Department of Statistics, North Carolina State University, Raleigh, North Carolina\\
\hskip 5mm \\

Wenyi Wang\\
Department of Statistics, North Carolina State University, Raleigh, North Carolina\\
\hskip 5mm \\

Yumeng Zhang\\
Department of Statistics, North Carolina State University, Raleigh, North Carolina\\
\hskip 5mm \\

%Ilsuk Kang \\
%Public Health Sciences Division, Fred Hutch Cancer Center, Seattle, Washington\\
%\hskip 5mm \\

Mike Lamonte\\
Department of Epidemiology and Environmental Health, University of Buffalo, Buffalo, NY\\
\hskip 5mm

Andrea LaCroix\\
Herbert Wertheim School of Public Health \& Human Longevity Science, University of California, San Diego, California\\
\hskip 5mm

Chongzhi Di\\
Public Health Sciences Division, Fred Hutchinson Cancer Center, Seattle, Washington\\

\end{center}

\begin{center}
{\Large{\bf Abstract}}\\
In large-scale epidemiological studies, it is often of interest to investigate joint relationships between longitudinal and time-to-event outcomes with exposures that are trajectories or functions. Our motivation study is the Objective Physical Activity and Cardiovascular Health (OPACH) Study, which collected accelerometry-measured physical activity in 6,489 older women. One of the scientific aims is to understand sedentary behavior accumulation patterns and its association with physical function (longitudinal) and mortality (time-to-event). We propose a novel approach that first converts raw accelerometry data into daily sitting bout accumulation profiles, which are treated as functional covariates, and then develop a functional joint model for longitudinal and time-to-event outcomes that incorporates a baseline functional covariate. The longitudinal process is modeled using functional data methods and linked to the survival process through functional principal component scores. Both sub-models include interpretable linear scalar-on-function regression coefficients to capture flexible dose-response associations between sitting accumulated across varying bout durations and health outcomes. Estimation is carried out via an efficient expectation–maximization (EM) algorithm with penalized spline approximations. Simulation studies demonstrate accurate parameter estimation and reliable model selection. Application to the OPACH data reveals flexible and interpretable dose-response relationships between sitting bout durations, physical function, and mortality. 
\end{center}
\baselineskip=12pt

\baselineskip=12pt
\par\vfill\noindent

\par\medskip\noindent
%\underline{\bf Short title}: Fast Multilevel Functional Principal Component Analysis

\clearpage\pagebreak\newpage
\pagenumbering{arabic}
\newlength{\gnat}
\setlength{\gnat}{22pt}
\baselineskip=\gnat

\section{Introduction} \label{sec:introduction}
Large-scale epidemiological studies increasingly collect exposures that are best represented
as functions or trajectories rather than scalar summaries, and interest often lies in relating
such functional exposures jointly to a longitudinally measured outcome and a time-to-event
outcome. This paper
develops a functional joint model that incorporates a baseline functional covariate into both
the longitudinal and survival sub-models, motivated by the Objective Physical Activity and
Cardiovascular Health (OPACH) study described below.

\subsection{The OPACH Study}

The Women’s Health Initiative (WHI) is a large prospective study designed to investigate major causes of morbidity and mortality among postmenopausal women. A total of 161,808 women aged 50–79 years were enrolled at 40 clinical centers across the United States between 1993 and 1998 \citep{women1998design}. 
%The WHI Long Life Study was a part of the second WHI extension study (2010-2015) in which in-home examinations were completed among a sub-cohort of 7,875. 
As an ancillary study of the WHI, the Objective Physical Activity and Cardiovascular Health (OPACH) Study aimed to investigate relationships between objectively measured physical activity and cardiovascular health and mortality, by collecting hip-worn accelerometry from 6,489 older women aged 63–99 years between 2012 and 2014 \citep{lacroix2017objective}.

%\subsection{Accelerometry-based sitting bout profiles}

Accelerometers are widely used in epidemiological studies to assess physical activity and sedentary behavior \citep{yang2010review}.
Device-specific software (e.g., ActiLife) typically converts raw acceleration signals into activity counts over pre-specified epochs using proprietary algorithms. A common approach for identifying sedentary behavior applies a threshold to epoch-level counts (e.g., $<$
100 counts per minute) and derives summary measures such as total sedentary time. 
%A sedentary bout is then defined as a continuous period of uninterrupted sedentary behavior.
However, threshold-based approaches may overestimate sitting time and inadequately characterize sitting patterns.  %{\color{red} Needs citation here}
%and can poorly estimate mean sitting bout duration when compared to measures obtained from criterion inclinometer devices (e.g., activPal) \citep{bellettiere2021agreement,carlson2019day,barreira2015free}. 
%is unique in that it utilizes two competitive architectures in the deep learning community, the convolutional neural network (CNN) and the bi-directional long short-term memory network (BiLSTM), for predicting sitting in 10-second epochs. 
After obtaining epoch-level sitting classifications, a natural next step is to characterize sitting bout (defined as a continuous period of uninterrupted sitting) accumulation patterns and examine their associations with health outcomes. Standard summary measures (e.g., mean or median bout duration) may fail to capture key features of these patterns.

To address these limitations, we propose to use a combination of deep learning and functional data methods. First, we adopt a recently developed convolutional neural network hip accelerometer posture (CHAP) algorithm to classify sitting status, which has demonstrated improved accuracy in identifying sitting at the 10-second epoch level \citep{greenwood2021cnn,eanes2018too}. Second, we characterize the full distribution of sitting bout durations viewed as functional data, instead of compressing the information into scalar summary measures. We refer to the sitting bout distributional data as a sitting bout accumulation profile.

\begin{figure}[h]
    \centering
   \includegraphics[width=0.75\textwidth]{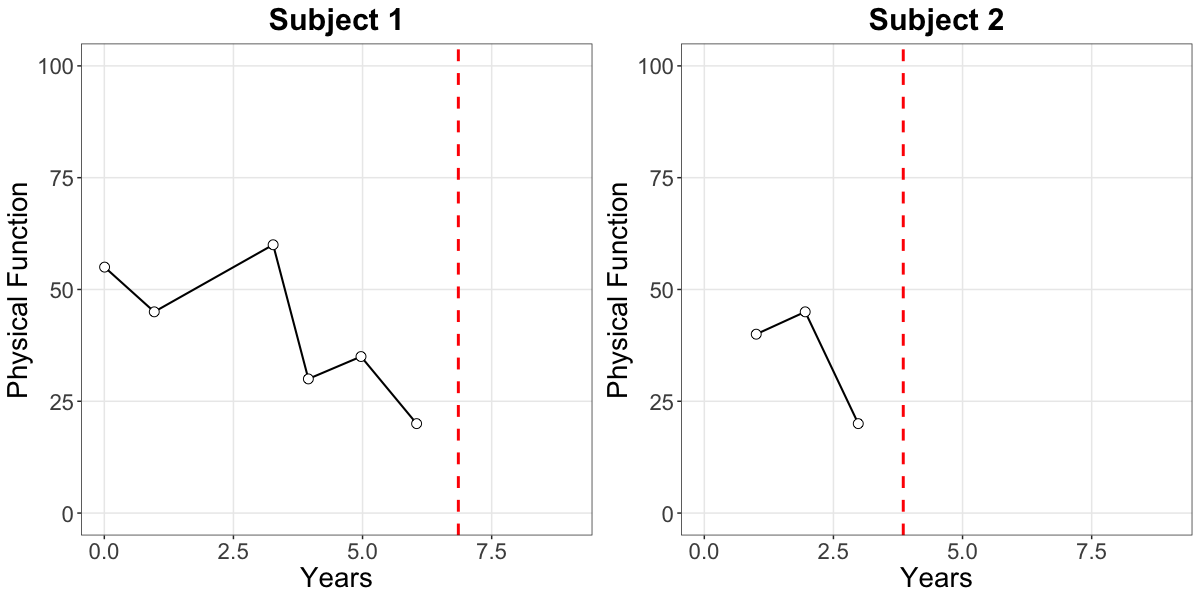}
    \caption{Observed SF-36 physical function score over time for the same two subjects as in Figure \ref{fig:intro:func_predictor}.  Each line connects repeated measurements of physical function score and red dashed vertical line indicates time of all-cause mortality.}
    \label{fig:intro:longi}
\end{figure}

Our objective is to assess the association between daily sitting profiles, longitudinal physical function scores obtained from surveys, and mortality. Figure~\ref{fig:intro:longi} displays the longitudinal trajectories of physical function scores for two representative subjects.  Subject 2 consistently exhibits lower physical function scores than Subject 1, consistent with the observation that Subject 2 accumulates more sitting time (Figure~\ref{fig:intro:func_predictor}).

\subsection{Existing literature on functional joint models}

Joint models (JM) are widely used for the joint analysis of longitudinal and survival data \citep{wulfsohn1997joint, tsiatis2004joint, 
he2024joint}. In recent years, functional joint models (FJM) have emerged, wherein longitudinal data are treated as sparse functional data \citep{yao2007functional, yan2017dynamic}. More advanced FJMs have been developed to accommodate multivariate sparse functional data \citep{li2022joint, zou2024aoas, wang2025multi}, recurrent event data \citep{hong2021dynamic}, and longitudinal images \citep{shi2024dynamic}.
Compared with parametric joint models, 
FJMs are generally more challenging to estimate, as they involve multiple nonparametric functions.
Consequently, many existing FJMs rely on two-stage estimation procedures for computational convenience, albeit at the cost of potential estimation bias. In contrast, \cite{li2022joint} proposed an expectation-maximization (EM) algorithm based on regression splines for estimating nonparametric functions, and \cite{wang2025multi} developed an EM algorithm using penalized splines \citep{Eilers1996}.

For clarity, we use the term functional joint modeling to refer specifically to settings in which the functional domain is longitudinal time, such as biomarkers measured at irregular time points. This should be distinguished from the extensive literature on Cox regression models with baseline functional covariates; see, for example, \cite{gellar2015cox}, \cite{qu2016optimal}, \cite{kong2018flcrm}, \cite{wang2020partial}, \cite{cui2021additive}, \cite{jiang2023predicting} and \cite{jiang2024functional}.

\subsection{Proposed method}

Although both functional joint models (FJM) and Cox models with functional covariates have been extensively studied, relatively few works have considered FJMs that additionally incorporate baseline functional covariates. Motivated by the OPACH data,  we propose a new functional joint model for the joint analysis of longitudinal outcomes-physical function scores collected from a survey- and a time-to-event outcome, namely all-cause mortality, while incorporating a baseline functional covariate consisting of daily sitting profiles obtained from accelerometers. As will be demonstrated later, both the longitudinal and survival sub-models include novel and interpretable linear scalar-on-function regression components to capture associations with the baseline functional covariate.

For the proposed model, we develop a computationally tractable expectation-maximization (EM) algorithm. Penalized splines \citep{Eilers1996} are employed to estimate the nonparametric smooth functions in the longitudinal sub-model, as well as the coefficient functions associated with the baseline functional covariate in both sub-models.
The penalized spline approach enables flexible modeling of nonlinear relationships while mitigating the overfitting risks commonly associated with regression splines. A key computational challenge lies in selecting smoothing parameters for multiple nonparametric functions. As noted in \cite{wang2025multi}, fast global selection algorithms for multiple smoothing parameters in nonparametric regression,  such as those developed for generalized additive models \citep{Wood2011},  are not directly applicable for FJMs. 

To address this issue, we exploit the iterative structure of the EM algorithms and adopt a local smoothing parameter selection strategy. 
Specifically, in the longitudinal sub-model, the estimation of smooth functions (including the mean function, eigenfunctions, and coefficient function) is reformulated as a weighted least squares problem for nonparametric regression and solved using the {\it mgcv} R package \citep{Wood2011,Wood2016}  at each EM iteration. This approach, previously developed in \cite{wang2025multi}, substantially improves the convergence of the EM algorithm.
In contrast,
selecting the smoothing parameter in the survival sub-model presents a new and more delicate challenge. We address this by developing a local model selection procedure based on the Akaike Information Criterion (AIC), which incorporates the expected survival log-likelihood along with a penalty term reflecting the complexity of the smooth coefficient function in the survival sub-model. %A predefined grid search identifies the optimal value of the smoothing parameter in each EM iteration. 

The remainder of the paper is organized as follows. Section~\ref{sec:profile} describes how raw accelerometry is transformed into daily sitting bout accumulation profiles. Section~\ref{sec:model} introduces the proposed functional joint model with a baseline functional covariate. Section~\ref{sec:estimate} describes the proposed Monte Carlo EM algorithm. Section~\ref{sec:model.selection} discusses the selection of key model parameters, including the number of functional principal components for modeling longitudinal data and the smoothing parameters. Section~\ref{sec:application} presents an application of the proposed method to the OPACH study. Section~\ref{sec:simulation} evaluates the model performance through simulation studies. Finally, Section~\ref{sec:discussion} concludes with a discussion of the findings and potential directions for future research.

% of implications, limitations, and future extensions. 
%The R code for implementing the proposed method is available at \href{https://github.com/wenyiwang2000/Multi-Cohort-FJM}{https://github.com/wenyiwang2000/Multi-Cohort-FJM}.

%\section{Daily sitting bout accumulation profiles}\label{sec:profile}

\section{Data processing: transforming raw accelerometry to interpretable sitting bout accumulation profiles}\label{sec:profile}
\begin{figure}[p]
    \centering
    \includegraphics[width=0.90\textwidth]{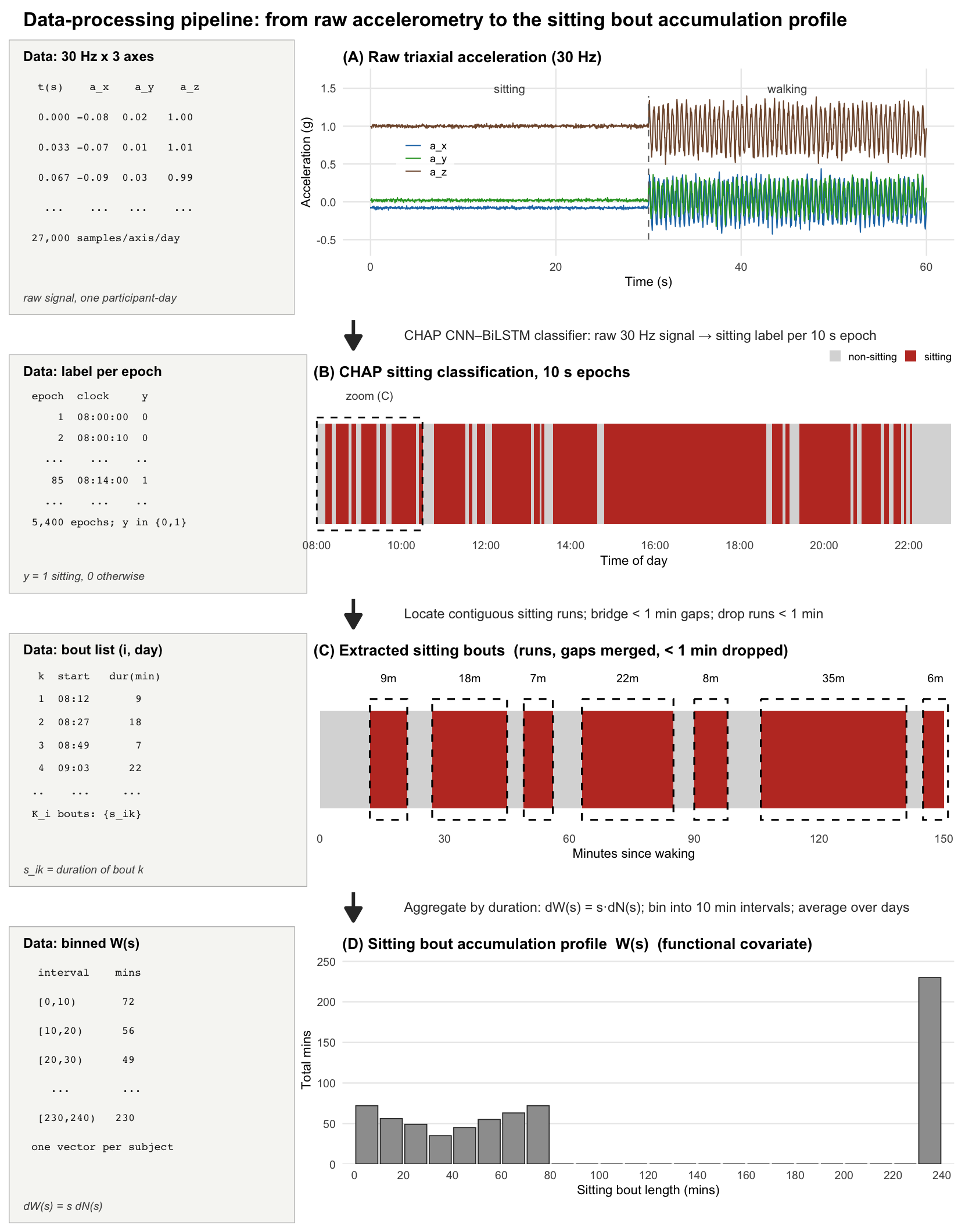}
    \caption{\footnotesize{Overview of the accelerometry data processing pipeline. Each stage shows both the data representation (left) and a visualization (right): (A) raw 30~Hz triaxial raw acceleration data; (B) CHAP sitting/non-sitting classification at the 10-second epoch level over one waking day (the dashed box marks the window magnified in panel~C); (C) sitting bouts extracted as contiguous sitting runs, with durations $s_{ik}$ annotated; and (D) the resulting accumulation profile $W(s)$ (shown in 10-minute bout-duration bins here).}}
    \label{fig:pipeline}
\end{figure}

In this section, we describe the processing of the accelerometry data. In the physical activity epidemiology literature, the standard procedure is to define sedentary behavior from minute-level activity counts using cut-points (e.g., $<100$ counts per minute) and then to compute simple summary metrics such as total sedentary minutes and mean sedentary bout duration. In contrast, we propose two main changes: (1) using the high-resolution (30~Hz) raw data, which provides much richer information, together with the deep-learning CHAP algorithm to predict sitting; and (2) using a functional data representation to characterize the full distribution of sitting bout durations for each participant, rather than a few summary metrics. These changes substantially improve both the accuracy and the richness of the resulting measurements. Relative to the standard count-based cut-point, CHAP classifies sitting far more accurately, especially at posture transitions (as detailed below), and the functional representation preserves the bout-duration information needed to reveal detailed dose-response association with health outcomes that simple summary measures cannot detect. Figure~\ref{fig:pipeline} provides an overview of the full pipeline, and the details of each step are described below.

\subsection{Applying deep learning algorithm to classify sitting}
The CHAP (Convolutional Neural Network Hip Accelerometer Posture) algorithm classifies sitting posture from hip-worn raw triaxial accelerometer data at the 10-second epoch level \citep{greenwood2021cnn}. It combines a convolutional neural network (CNN) for spatial feature extraction with a bidirectional long short-term memory network for temporal modeling, which substantially improves classification at posture transitions between sitting, standing, and walking. CHAP was trained and validated in 709 older adults who concurrently wore a hip-mounted ActiGraph GT3X+ and a thigh-worn activPAL inclinometer, the criterion standard for posture detection \citep{greenwood2021cnn}. Against the activPAL labels, CHAP achieved 93\% minute-level agreement and 83\% positive predictive value (PPV) for sitting-to-stand transitions, compared with 73\% agreement and only 30\% PPV for the conventional cut-point method. Thus, CHAP provides accurate, high-resolution sitting classification from hip-worn accelerometers without requiring thigh-mounted devices in large-scale studies.

For the OPACH study, we applied the CHAP algorithm to raw 30 Hz accelerometry data collected from hip-worn monitors worn by $n = 6{,}489$ women (aged 63--99 years) over 7 consecutive days. The CHAP model processes each day's raw 30 Hz triaxial acceleration signal and outputs a binary sitting/non-sitting label in 10-second epochs. These epoch-level classifications serve as the input to the subsequent data processing steps described in Section~\ref{sec:profile} below. {After excluding participants with insufficient valid wear days or incomplete covariate and outcome data, $5{,}706$ women were retained for the analysis in Section~\ref{sec:application}.

\subsection{Characterizing sitting accumulation patterns using a functional data framework}

We construct a daily sitting bout accumulation profile for each subject through {three steps, building on the epoch-level sitting classification of Section~\ref{sec:profile} (Step~1): bout extraction (Step~2) and aggregation into a functional representation (Step~3)}.
In Step 2, we identify individual sitting bouts from the binary sequence of 10-second sitting/non-sitting epochs by locating continuous sitting epochs and defining bout duration as the total elapsed time from the {start of the first to the end of the last} consecutive sitting epoch in each run. 
%A minimum bout duration threshold (e.g., 1 minute) is applied to exclude trivial episodes such as brief postural readjustments, while very short inter-episode gaps shorter than this threshold are bridged so that isolated non-sitting epochs within an otherwise continuous sitting period do not artificially fragment sustained sitting episodes. 
From the epoch-level sequence, we record the start time and duration of each bout, yielding a list $\{s_{ik}\}_{k=1}^{K_i}$ of $K_i$ bouts with durations expressed in minutes for subject $i$ {on a given day; we suppress the day index $j$ here and reinstate it in the multi-day formulation of Appendix~A}.

We formalize Step 3 using counting process notation. For simplicity, consider a single day of accelerometry data for subject $i$. Let $K_i$ denote the  number of sitting bouts,  and let $s_{ik} (k=1,\ldots, K_i )$ be the duration of the $k$th bout. 
Let $\mathcal{S}$ be a compact interval containing all possible bout durations, and define
$N_i(s)=\sum_{k=1}^{K_i} 1_{\{s_{ik} \leq s\}},\, s \in \mathcal{S}$, which counts the number of sitting bouts with duration less than or equal to $s$. Then $s\,\mathrm{d}N_i(s)$ represents the total time contributed by bouts of duration $s$. For example, three sitting bouts of $30$ minutes contribute 90 minutes, while a single bout of $60$ minutes contributes  60 minutes. Thus, total sitting time is obtained by summing contributions across
bouts of the same duration. For multiple days of accelerometry data,  we average $s \,\mathrm{d} N_i(s)$ across days;  details are provided in Appendix A.

Figure~\ref{fig:intro:func_predictor} displays  sitting bout accumulation profiles for two subjects, illustrating the distribution of total sitting times across sitting bout durations. {Because durations are recorded in 10-second multiples, they are effectively discrete; for visualization only, we bin them into intervals (in minutes):} $[0, 10)$, $[10, 20),...$, and $[240, \infty)$. {The estimation in Sections~\ref{sec:model}--\ref{sec:model.selection} uses the exact bout durations $s_{ik}$ and does not rely on this binning.}
For example,  Subject 1 accumulates approximately 75 minutes per day in bouts shorter than 10 minutes, whereas Subject 2 accumulates approximately 125 minutes. {The plots show that Subject 2 has longer total sitting time and more frequent short bouts, whereas Subject 1 has a single long bout ($\geq 240$ minutes).} The plotted totals are averages over multiple days (five days in this example), which explains the scale of the $y$-axis. 

%Figure~\ref{fig:intro:func_predictor} displays  sitting bout accumulation profiles for two subjects, illustrating the distribution of total sitting times across sitting bout durations. Because duration are not observed continuously, we discretize time into intervals (in minutes): $[0, 10)$, $[10, 20),...$, and $[240, \infty)$.
%For example,  Subject 1 accumulates approximately 75 minutes per day in bouts shorter than 10 minutes, whereas Subject 2 accumulates approximately 125 minutes. The plots clearly show that Subject 2 had longer sitting times, with more frequent short duration sitting bouts. Subject 2 exhibits more frequent short bouts, while Subject 1 has a long bout ($\geq 240$ minutes). The plotted totals are averages over multiple days (five days in this example), which explains the scale of the $y$-axis.

%Noticeably, subject 2 has not only a longer total sitting time, but also longer sitting bouts compared to subject 1, which agrees with the subject's much lower values of physical function score compared to those of subject 1.

\begin{figure}[h]
    \centering
   \includegraphics[width=0.75\textwidth]{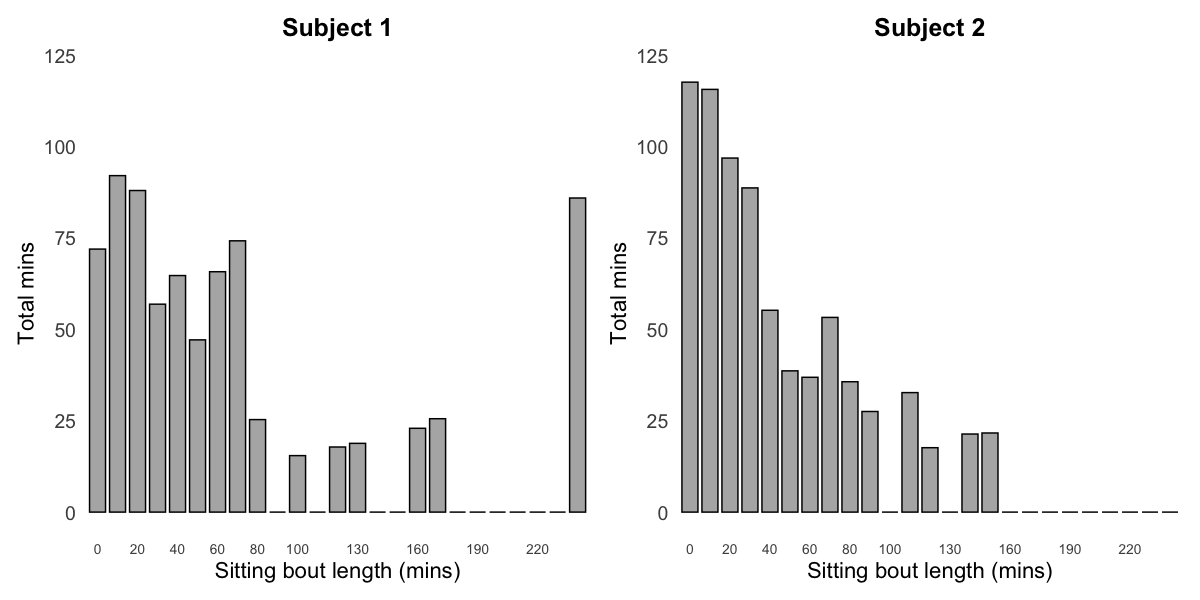}
    \caption{Daily sitting bout accumulation profiles for two subjects. The bars in both plots show the total sitting times of sitting bouts of length in each time interval by minute: $[0,10), [10, 20),\ldots, [240, +\infty)$ and averaged across multiple days.}
    \label{fig:intro:func_predictor}
\end{figure}

To motivate the model developed in Section~\ref{sec:model}, we briefly preview how the sitting profile enters the regression. We  treat sitting bout accumulation profiles as functional data, enabling a flexible and interpretable representation of the association between accelerometer-derived data and health outcomes.
Let $W_i(s) = s\, \mathrm{d} N_i(s)$ (see Appendix A for the multi-day case). We model the association via
 $\int_{s\in \mathcal{S}}\beta(s) \mathrm{d} W_i(s)=\int_{s\in \mathcal{S}}\beta(s) s\,\mathrm{d}N_i(s)$, where $\beta(s)$ is a smooth coefficient function.
 If $\beta(s)=c$, a constant, then the association reduces to $\int_{s\in \mathcal{S}}\beta(s)s\mathrm{d}N_i(s)=c(\sum_{k=1}^{K_i}s_{ik})$, i.e., it depends only on total sitting time.
If $\beta(s)$ varies with $s$, the duration of bouts influences the strength of association beyond total sitting time.
For example, a decreasing 
$\beta(s)$ implies that longer bouts have a stronger (more negative) association than shorter bouts with the same total duration. This functional formulation allows the data to reveal duration-specific effects. In the OPACH application, the estimated non-constant 
$\beta(s)$ provides evidence that bout duration  affects associations with both physical function and all-cause mortality—patterns that would not be captured by models based solely on total sitting time.

\section{Functional Joint Model with a Daily Sitting Bout Accumulation Profile} \label{sec:model}

\subsection{Data structure and notation}

Let $n$ be the number of subjects, indexed by $i$. For subject $i$, let $Y_{ij}$ be the $j$th  physical function score  from the SF-36 survey measured at  time $t_{ij}$, with $1\leq j \leq m_{i}$, where $m_{i}$ is the number of observations. The observational times satisfy $t_{ij}\in\mathcal{T}$, a compact interval representing the follow-up period. Let $S_i$ denote the the time from baseline to call-cause death. When $S_i$ is right-censored, we observe $T_i = \text{min}(S_i,C_i)$, where $C_i$ is the censoring time, assumed independent of both  $S_i$ and the longitudinal observations. For the OPACH data, $C_i$ corresponds to the last follow-up time and is non-informative.
Let $\Delta_i = 1_{\{S_i\leq C_i\}}$ be the event indicator. Longitudinal measurements are observed only up to $T_i$, i.e., $t_{ij} \in [0, T_i] \subset \mathcal{T}$. Let $\boldsymbol{Z}_i = (Z_{i1}, \ldots, Z_{iP}) \in \mathbb{R}^P$ denote baseline scalar covariates, and let $W_i(s), s \in \mathcal{S}$ denote the daily cumulative sitting time for bouts of duration not exceeding $s$.% functional covariate, where $\mathcal{S}$ is a compact interval.

\subsection{Longitudinal model with a daily sitting bout accumulation profile}\label{sec:FLFM}

We model the longitudinal outcome as sparse functional data and consider the following functional linear regression model
\begin{equation}
\begin{split}
    &Y_{ij} = \mu(t_{ij}) + \bZ_i^{\top}\bgamma_1 + \int_{s\in \mathcal{S}}\beta_1(s)\mathrm{d}W_i(s) + X_{i}(t_{ij}) + \epsilon_{ij},
\end{split}
\label{eq:longitudinal}
\end{equation}
where $\mu(t)$ is the fixed smooth intercept function of $t$, $\bgamma_1$ is the coefficient vector for $\bZ_i$, and $\beta_1(s)$ is an unknown smooth coefficient function associated with the sitting  profile $W_i(s)$. The subject-specific process
$X_i(t)$ captures deviations from the mean, and $\epsilon_{ij}$ denotes measurement error. We assume  $\bepsilon_i = (\epsilon_{i1},\ldots, \epsilon_{im_i})^\top$ is   independent of $X_i(t)$.

We model  $X_i(t)$ as a zero-mean Gaussian process with covariance function $\mathcal{C}(t, t') = \text{Cov}\{X_{i}(t), X_{i}(t')\}$. We assume that the covariance function is of low rank so that 
$\mathcal{C}(t, t') = \sum_{\ell=1}^L \lambda_{ \ell} \phi_{\ell}(t) \phi_{\ell}(t')$, where $\lambda_{1}\geq \lambda_{2}\geq\ldots \geq \lambda_L> 0$ are eigenvalues and $\phi_{\ell}(t)$ are  orthonormal eigenfunctions on $\mathcal{T}$, satisfying $\int_{\mathcal{T}} \phi_{\ell}(t) \phi_{ \ell'}(t) \mathrm{d}t = 1_{\{\ell=\ell'\}}$. The eigenfunctions $\phi_\ell(t)$ characterize the varying patterns of the longitudinal outcome; as there are only a few observations for each subject, it is reasonable to assume that there are only a small number ($L$) of varying patterns in modeling $X_i(t)$.
By the Karhunen-Loeve expansion,  $X_{i}(t) = \sum_{\ell=1}^{L}\xi_{i \ell}\phi_{\ell}(t)$ with random functional principal component (FPC) scores $\xi_{i\ell} \sim N(0,\lambda_{\ell})$. Let $\boldsymbol{\xi}_i = (\xi_{i1},\ldots, \xi_{iL})^{\top}$ with  covariance  $\boldsymbol{\Lambda}=\text{diag}(\lambda_{1},\ldots, \lambda_{L})$. 
The number $L$ will be  selected via AIC/BIC. Measurement errors $\bepsilon_{i}$ are assumed to be a multivariate normal with zero mean and covariance $\sigma^2 \boldsymbol{I}_{m_i}$. The independence between the stochastic process $X_i(t)$ and the random errors implies that $\bxi_i$ and $\bepsilon_i$ are independent.
%Given the eigenfunctions $\phi_{\ell}(t)$s, Model~\eqref{eq:longitudinal} has the mixed model representation:
%\begin{equation*}\label{eq:mixed}
%Y_{ij} = \mu(t_{ij}) + \bZ_i^{\top}\bgamma_1 + \int_{s\in %\mathcal{S}}W_i(s)\beta_1(s)\mathrm{d}s + \sum_{\ell=1}^L\xi_{i %\ell}\phi_{\ell}(t_{ij}) + \epsilon_{ij}.
%\end{equation*}

\subsection{Joint model for longitudinal and survival data with a daily sitting bout accumulation profile}
Let $h_i(t)$ denote the hazard function for subject $i$. We extend  the Cox proportional hazards model to incorporate the daily sitting bout accumulation profile as a baseline functional covariate,
\begin{equation}\label{eq:hazard}
    h_i(t) = h_0(t) \text{exp}\left\{\bZ_i^{\top}\boldsymbol{\gamma}_{2} + \int_{s \in \mathcal{S}}  \beta_2(s)\mathrm{d}W_i(s) + \boldsymbol{\xi}^{\top}_{i}\bgamma_3\right\}\;,
\end{equation}
where $h_0(t)$ is the baseline hazard function,  $\bgamma_2$ is the coefficient vector for  $\bZ_i$, $\beta_2(s)$ is an unknown smooth coefficient function for the sitting profile $W_i(s)$, and $\bgamma_3=(\gamma_{31},\ldots,\gamma_{3L})^{\top}$ is the coefficient vector associated with the score vector $\bxi_i$, linking longitudinal and survival processes.
If $\bgamma_3=0$, the two processes are conditionally independent.

\subsection{Likelihood of joint model}

Let $\boldsymbol{y}_{i} = (Y_{i1}, \ldots,Y_{im_{i}})^{\top}$, $\bmu_i=(\mu(t_{i1}), \ldots,\mu(t_{im_{i}}))^{\top}$, and $\boldsymbol{F}_i=\Big(\bZ_i^{\top}\bgamma_1 + \int_{s\in \mathcal{S}}\beta_1(s)\mathrm{d}W_i(s) \Big)\boldsymbol{1}_{m_i}$. % be the vectors of longitudinal observations, the intercept function evaluated at the observed time points, and the effect of the baseline covariates for the $i$th subject, respectively. 
Define $\boldsymbol{\phi}(t) =\{\phi_1(t),\ldots, \phi_L(t)\}^\top$  and $\boldsymbol{\Phi}_{i} = [\boldsymbol{\phi}(t_{i1}),\cdots, \boldsymbol{\phi}(t_{im_i})]^\top$. The latent trajectory is  $\boldsymbol{x}_{i} =\boldsymbol{\mu}_i + \boldsymbol{F}_i + \boldsymbol{\Phi}_{i}\boldsymbol{\xi}_{i}$ with error variance $\boldsymbol{\Sigma}_i = \sigma^2\boldsymbol{I}_{m_{i}}$. The conditional likelihood of the longitudinal data given $\bxi_i$ is
\begin{equation*}%\label{eq:ml_likelihood}
    f(\boldsymbol{y}_{i}|\boldsymbol{x}_{i},\boldsymbol{\Sigma}_i) = (|2\pi \boldsymbol{\Sigma}_i|)^{-\frac{1}{2}}\text{exp}\left\{ -\frac{1}{2}(\boldsymbol{y}_{i}-\boldsymbol{x}_{i})^{\top}\boldsymbol{\Sigma}_i^{-1}(\boldsymbol{y}_{i}-\boldsymbol{x}_{i})
    \right\}.
\end{equation*}
The distribution of random scores $\boldsymbol{\xi}_{i}$ is $f(\boldsymbol{\xi}_{i}|\boldsymbol{\Lambda}) = (|2\pi \boldsymbol{\Lambda}|)^{-\frac{1}{2}} \text{exp}(-\frac{1}{2}\boldsymbol{\xi}_{i}^{\top}\boldsymbol{\Lambda}^{-1}\boldsymbol{\xi}_{i})$. %The joint distribution $f(\boldsymbol{y}_{i},\boldsymbol{\eta}_{i}|\boldsymbol{\Lambda}_{1},\boldsymbol{\Lambda}_{2})$ is shown in Web Appendix~\ref{app:joint}.

The conditional partial likelihood of the survival outcome is
\begin{equation}\label{eq:survival}
\begin{split}
    &f(T_i,\Delta_i|h_0, \boldsymbol{Z}_i,W_i(s),\boldsymbol{\xi}_i, \boldsymbol{\gamma}_{2},\beta_2(s),\boldsymbol{\gamma}_{3})\\
    = &\left\{h_0(T_i)\exp\left(\bZ_i^{\top}\boldsymbol{\gamma}_{2} + \int_{s \in \mathcal{S}} \beta_2(s)\mathrm{d}W_i(s)  + \boldsymbol{\xi}^{\top}_{i}\bgamma_3\right)\right\}^{\Delta_i}\\
     &\times\text{exp}\left\{-\Lambda_0(T_i)\exp\left( \bZ_i^{\top}\boldsymbol{\gamma}_{2} + \int_{s \in \mathcal{S}}  \beta_2(s)\mathrm{d}W_i(s) + \boldsymbol{\xi}^{\top}_{i}\bgamma_3\right)\right\},
\end{split}
\end{equation}
where $\Lambda_0(t) = \int_0^t h_0(u)\mathrm{d}u$ is the baseline cumulative hazard function. 

%The longitudinal data and the time-to-event data are assumed to be conditionally independent given the random effects $\boldsymbol{\xi}_i$. This assumption reflects the idea that the subject-specific trajectory $\boldsymbol{\xi}_{i}$ sufficiently capture the dependence structure between the longitudinal and survival components. 

%The overall marginal likelihood is obtained by integrating the product of the longitudinal and survival likelihoods over the distributions of the random score vector $\boldsymbol{\xi}_i$. 
Finally, the  marginal log-likelihood is obtained by integrating out $\bxi_i$:
\begin{equation}\label{eq:marginal}
\sum_{i=1}^n \log \left\{\int f(\boldsymbol{y}_i|\boldsymbol{x}_i,\boldsymbol{\Sigma}_i) f(T_i,\Delta_i|h_0, \boldsymbol{Z}_i,W_i(s),\boldsymbol{\xi}_i, \boldsymbol{\gamma}_{2},\beta_2(s),\boldsymbol{\gamma}_{3})
f(\boldsymbol{\xi}_{i}|\boldsymbol{\Lambda})\mathrm{d}\boldsymbol{\xi}_i\right\}.
\end{equation}

\section{Model Estimation via Monte Carlo EM}\label{sec:estimate}

\subsection{Spline approximation of smooth functions} 

The smooth intercept function, eigenfunctions, and coefficient functions in the longitudinal model are approximated by splines; so is the coefficient function in the survival model.
Let $\boldsymbol{b}(t)=\{B_1(t),\cdots,B_{K}(t)\}^{\top}$ be the vector of the B-spline basis functions on $\mathcal{T}$ and $\boldsymbol{b}^{\beta}(s)=\{B_1^{\beta}(s),\cdots,B_{K^{\beta}}^{\beta}(s)\}^{\top}$ be the vector of the B-spline basis functions on $\mathcal{S}$, where $K$ and $K^{\beta}$ are the numbers of equally-spaced interior knots plus the order (degree + 1) of the B-splines in $\mathcal{T}$ and $\mathcal{S}$, respectively. The intercept function $\mu(t)$ is modeled as $\mu(t)=\boldsymbol{b}(t)^{\top}\boldsymbol{\alpha}$, where $\boldsymbol{\alpha}$ is the corresponding coefficient vector. The smooth coefficient functions for the baseline functional covariate $W_i(s)$ in the longitudinal model and the survival model are approximated as $\beta_1(s)=\boldsymbol{b}^{\beta}(s)^{\top}\boldsymbol{\omega}_1$ and $\beta_2(s)=\boldsymbol{b}^{\beta}(s)^{\top}\boldsymbol{\omega}_2$, where $\bomega_1$ and $\bomega_2$ are the corresponding coefficient vectors.   

To ensure numerical stability, we transform the bases of the B-spline in $\mathcal{T}$ using the Gram matrix: $\boldsymbol{G} = \int \boldsymbol{b}(t)\boldsymbol{b}(t)^{\top}dt \in \mathbb{R}^{K\times K}$. Then the transformed B-spline bases are  $\widetilde{\boldsymbol{b}}(t) = \boldsymbol{G}^{-1/2}\boldsymbol{b}(t)$ and they are orthonormal. %This orthonormalization ensures numerical stability by mitigating issues such as multicollinearity and ill-conditioned matrix inversions, which can arise in high-dimensional functional data settings. 
The eigenfunctions of the
 covariance function $\mathcal{C}(t, t')$ are then approximated by $\phi_{\ell}(t)=\widetilde{\boldsymbol{b}}(t)^{\top}\boldsymbol{\theta}_{\ell}$, where $\boldsymbol{\theta}_{\ell}$ is the coefficient vector for the $\ell$-th eigenfunction. The orthonormality of the eigenfunctions leads to the  constraints: $\boldsymbol{\theta}_{\ell_1}^{\top}\boldsymbol{\theta}_{\ell_2} = 1_{\{\ell_1= \ell_2\}}$. These orthonormality constraints ensure that the eigenfunctions are linearly independent and form a valid basis for the decomposition of the covariance function $\mathcal{C}(t, t')$. %This finite-dimensional approximation reduces model complexity while retaining flexibility, facilitating efficient parameter estimation in the subsequent EM algorithm steps.

\subsection{E-step}
The spline approximation of nonparametric smooth functions enables the use of parametric estimation methods for the functional joint model. Direct maximization of the marginal log-likelihood in Model~\eqref{eq:marginal} is computationally challenging; therefore, we employ an expectation-maximization (EM) algorithm \citep{Dempster1977}, treating the FPC score vector $\boldsymbol{\xi}_i$ as latent  data. %The EM algorithm iteratively alternates between two steps until convergence: (i) computing the expectation of the joint log-likelihood of the observed data with respect to the score vector given the current parameter estimates (E-step) and (ii) maximizing the expected joint log-likelihood to update the parameter estimates (M-step).

Let $\mathbb{Y}_i = \{\boldsymbol{y}_{i}, \boldsymbol{Z}_{i}, W_i(s),T_i, \Delta_i\}$ denote the observed data for subject $i$, and let $\boldsymbol{\Theta} = [\boldsymbol{\theta}_{1},\cdots,\boldsymbol{\theta}_{ L}]$.  Denote the  full  parameter set by  $\boldsymbol{\Omega} = \{h_0, \boldsymbol{\alpha}, \bgamma_1, \boldsymbol{\omega}_1, {\sigma}^2,\boldsymbol{\gamma}_2,\bomega_2,\bgamma_3,\boldsymbol{\Lambda},\boldsymbol{\Theta}\}$. The E-step computes the conditional expectation of  the complete-data joint log-likelihood given the observed data and current parameter estimates, $\boldsymbol{\widehat{\Omega}}$:
\begin{equation}\label{eq:expectation}
\begin{split}
Q(\boldsymbol{\Omega}) = &\sum_{i=1}^n\mathbb{E}_i \Big\{\log\big(f(\boldsymbol{y}_i|\boldsymbol{x}_i,\boldsymbol{\Sigma}_i)\big)\Big\} + 
\sum_{i=1}^n\mathbb{E}_i \Big\{\log\big(f(\boldsymbol{\xi}_{i}|\boldsymbol{\Lambda})\big)\Big\}
\\
& + \sum_{i=1}^n\mathbb{E}_i \Big\{\log\big(f(T_i,\Delta_i|h_0, \boldsymbol{Z}_i,W_i(s),\boldsymbol{\xi}_i, \boldsymbol{\gamma}_{2},\bomega_2,\boldsymbol{\gamma}_{3})\big)\Big\},
\end{split}
\end{equation}
where $\mathbb{E}_i(\cdot)$ denotes expectation with respect to the conditional distribution of $ \boldsymbol{\xi}_i|\mathbb{Y}_i, \boldsymbol{\widehat{\Omega}}$. 

For a smooth function $g(\cdot)$, the conditional expectation $\mathbb{E}_i\{g(\boldsymbol{\xi}_i)\}$ can be expressed as
\begin{equation*}
 \frac{\int g(\boldsymbol{\xi}_i)f(T_i,\Delta_i|\boldsymbol{Z}_i, W_i(s), \boldsymbol{\xi}_i, \boldsymbol{\widehat{\Omega}})f(\boldsymbol{\xi}_i|\boldsymbol{y}_i,\boldsymbol{Z}_i, W_i(s),\boldsymbol{\widehat{\Omega}})\mathrm{d}\boldsymbol{\xi}_i}{\int f(T_i,\Delta_i|\boldsymbol{Z}_i, W_i(s), \boldsymbol{\xi}_i, \boldsymbol{\widehat{\Omega}})f(\boldsymbol{\xi}_i|\boldsymbol{y}_i,\boldsymbol{Z}_i, W_i(s),\boldsymbol{\widehat{\Omega}})\mathrm{d}\boldsymbol{\xi}_i}.
 \end{equation*}
Here, $f(T_i,\Delta_i|\boldsymbol{Z}_i, W_i(s), \boldsymbol{\xi}_i, \boldsymbol{\widehat{\Omega}})$  is multivariate normal (see Web Appendix B). 

We approximate these expectations using Monte Carlo integration. Specifically, we draw $R$ samples $\boldsymbol{\xi}^{(r)}_i$ from the multivariate normal distribution $f(\boldsymbol{\xi}_i|\boldsymbol{y}_i, \boldsymbol{Z}_i, W_i(s),\boldsymbol{\widehat{\Omega}})$ and compute:
\begin{equation*}
    \mathbb{E}_i\big\{g(\boldsymbol{\xi}_i)\big\} \approx \frac{\sum_{r=1}^R g(\boldsymbol{\xi}^{(r)}_i)f(T_i,\Delta_i|\boldsymbol{Z}_i, W_i(s), \boldsymbol{\xi}_i^{(r)}, \boldsymbol{\widehat{\Omega}})}{\sum_{r=1}^Rf(T_i,\Delta_i|\boldsymbol{Z}_i, W_i(s), \boldsymbol{\xi}^{(r)}_i,\boldsymbol{\widehat{\Omega}})}.
\end{equation*}
%The Monte Carlo approximation enables efficient implementation of the E-step, ensuring computational feasibility even in high-dimensional parameter spaces. 

\subsection{M-step} 
In the M-step, the parameter vector $\boldsymbol{\Omega}$ is updated by maximizing $Q(\boldsymbol{\Omega})$ in~(\ref{eq:expectation}). 
The updates are carried out separately for the longitudinal parameters, survival parameters, and eigenvalues.

%Specifically, estimates of longitudinal parameters $\big\{\boldsymbol{\alpha}, \boldsymbol{\gamma}_1, \bomega_1, \boldsymbol{\sigma}^2, \boldsymbol{\Theta}\big\}$, survival parameters $\big\{h_0,\boldsymbol{\gamma}_2, \bomega_2,\bgamma_3 \big\}$, and eigenvalue parameters in $\boldsymbol{\Lambda}$ are updated iteratively. In the following, we provide details on the estimation process for each set of parameters.

First, the longitudinal parameters $\big\{\boldsymbol{\alpha}, \boldsymbol{\gamma}_1, \bomega_1, \boldsymbol{\sigma}^2, \boldsymbol{\Theta}\big\}$ are updated  by minimizing the expected negative log-likelihood of the longitudinal model:
\begin{equation}\label{eq:mini_long}
\sum_{i=1}^n\Bigg\{ \frac{m_{i}}{2}\text{log}(2\pi \sigma^2) + \frac{1}{2\sigma^2}\mathbb{E}_i\Big\Vert\boldsymbol{y}_{i} - \boldsymbol{B}_{i}\boldsymbol{\alpha}- \bZ_i^{\top}\bgamma_1\boldsymbol{1}_{m_i}- \Big(\int_{s\in \mathcal{S}}\boldsymbol{b}^{\beta}(s)^{\top}\bomega_1 \mathrm{d}W_i(s)\Big) \boldsymbol{1}_{m_i} - \widetilde{\boldsymbol{B}}_{i}\boldsymbol{\Theta}\boldsymbol{\xi}_{i}\Big\Vert^2\Bigg\},
\end{equation}
where $\Vert\cdot\Vert$ is the Euclidean norm, $\boldsymbol{B}_{i} =[\boldsymbol{b}(t_{i1}),\ldots, \boldsymbol{b}(t_{im_{i}})]^{\top}$ is the matrix of the B-spline bases evaluated at the time points $t_{i1}, \ldots, t_{im_{i}}$, and $\widetilde{\boldsymbol{B}}_{i} = \boldsymbol{G}^{-1/2}\boldsymbol{B}_{i}$. To enforce smoothness of the mean function, the coefficient function and the eigenfunctions, we adopt penalized spline estimation \citep{Eilers1996}. Compared with  regression spline estimation, penalized spline estimation improves numerical stability and yields smoother and more interpretable estimates.

However, selection of smoothing parameters is challenging in the EM framework. We address this by adopting a local selection strategy, taking advantage of  the iterative structure of EM algorithms. 
At each iteration, estimation of the smooth functions is reformulated as a weighted least squares problem and solved using the {\it mgcv} R package \citep{Wood2011,Wood2016}, which 
simultaneously estimates smoothing parameters. Integrals of the form
$\int_{s\in \mathcal{S}}\boldsymbol{b}^{\beta}(s)^{\top}\bomega_1 \mathrm{d}W_i(s)$
are computed as: $\sum_{k=1}^{K_i} s_{ik} \boldsymbol{b}^{\beta}(s_{ik})^{\top}\bomega_1$.
%by $\frac{1}{J-1}\sum_{j_s=1}^{J-1} \{ W_i(s_{j_s+1}) - W_i(s_{j_s})\} \boldsymbol{b}^{\beta}(s_{j_s})^{\top}\bomega_1$, where $\{s_1, \ldots, s_{J}\}$ is a regular and dense grid on $[0,1]$.
In practice, we use seven B-spline bases constructed at equally spaced knots on both $\mathcal{T}$ and $\mathcal{S}$ ($K=K^{\beta}=7$) for simulations and real data. Orthonormality of the  eigenfunctions is  enforced via post-processing. Details and the observed Fisher information matrix for the model parameters are provided in Web Appendix D.

Second, the parameters $\big\{h_0,\boldsymbol{\gamma}_2, \bomega_2,\bgamma_3 \big\}$ in the Cox regression are updated by minimizing the expected negative survival log-likelihood:
\begin{equation*}\label{eq:mini_surv}
\begin{split}
    \sum_{i=1}^{n}\Bigg\{&-\Delta_i\Big\{\text{log}(h_0(T_i)) +  \bZ_i^{\top}\boldsymbol{\gamma}_{2} + \int_{s \in \mathcal{S}}  \boldsymbol{b}^{\beta}(s)^{\top}\bomega_2\mathrm{d}W_i(s) + \mathbb{E}_i\big(\boldsymbol{\xi}_{i}^{\top}\boldsymbol{\gamma}_{3}\big) \Big\} + \\&\Lambda_0(T_i)\mathbb{E}_i\Big\{\exp\big(\bZ_i^{\top}\boldsymbol{\gamma}_{2} + \int_{s \in \mathcal{S}}  \boldsymbol{b}^{\beta}(s)^{\top}\bomega_2\mathrm{d}W_i(s) + \boldsymbol{\xi}_{i}^{\top}\boldsymbol{\gamma}_{3} \big)\Big\} \Bigg\}.
\end{split}
\end{equation*}
%where
%$\int_{s \in \mathcal{S}} \boldsymbol{b}^{\beta}(s)\mathrm{d}W_i(s) =(\int_{s \in \mathcal{S}}  B^{\beta}_1(s)\mathrm{d}W_i(s), \ldots, \int_{s \in \mathcal{S}}  B^{\beta}_{K^{\beta}}(s)\mathrm{d}W_i(s))^{\top}$. 
The baseline hazard $h_0(t)$ is updated using a Breslow-type estimator:
\begin{equation*}
  \widehat{h}_0(t) = \frac{\sum_{i=1}^{n}\Delta_i{1}_{\{T_i=t\}}}{\sum_{i=1}^{n}\mathbb{E}_i\Big\{\text{exp}\big(\boldsymbol{Z}_{i}^{\top}\boldsymbol{\gamma}_{2} + 
  \int_{s \in \mathcal{S}} \boldsymbol{b}^{\beta}(s)^{\top}\bomega_2 \mathrm{d}W_i(s)  +\boldsymbol{\xi}_{i}^{\top}\boldsymbol{\gamma}_{3}\big)\Big\}{1}_{\{T_i\geq t\}}} \;.
\end{equation*}
To ensure smoothness of $\beta_2(s)$, we include a qudaratic penalty
$\tau_2^{\beta}\bomega_2^{\top}\boldsymbol{P}^{\beta}\bomega_2$,
where $\boldsymbol{P}^{\beta}$ is a penalty matrix and $\tau_2^{\beta}$ is the smoothing parameter. The parameters are updated using a Newton-Raphson algorithm:
$
\widehat{\boldsymbol{\gamma}}_{d} = \widehat{\boldsymbol{\gamma}}_{d-1} + \boldsymbol{I}^{-1}_{\widehat{\boldsymbol{\gamma}}_{d-1}}\boldsymbol{S}_{\widehat{\boldsymbol{\gamma}}_{d-1}},
$
where $\boldsymbol{S}_{\widehat{\boldsymbol{\gamma}}_{d-1}}$ and $\boldsymbol{I}_{\widehat{\boldsymbol{\gamma}}_{d-1}}$ denote
the  penalized score vector and observed information matrix, respectively. Explicit expressions are provided in Web Appendix D.

Third, the eigenvalues $\{\lambda_\ell\}_{\ell=1}^L$ are updated by minimizing $-\sum_{i=1}^{n}\mathbb{E}_i \{\log(f(\boldsymbol{\xi}_{i}|\boldsymbol{\Lambda}))\}$,
% This minimization simplifies to $\sum_{i=1}^n \Big\{\log|\boldsymbol{\Lambda}|+\mathbb{E}_i(\boldsymbol{\xi}^T_{i}\boldsymbol{\Lambda}^{-1}\boldsymbol{\xi}_{i})\Big\}$. The estimator for the eigenvalues is 
yielding the closed-form estimator
$\widehat{\lambda}_{\ell}= n^{-1} \sum^n_{i=1}\mathbb{E}_i({\xi}^2_{i\ell})$, $\ell=1,\ldots, L$.

\section{Model Selection} \label{sec:model.selection}
Penalized splines \citep{Eilers1996} are used to estimate the intercept function and eigenfunctions in the longitudinal data model, as well as the coefficient functions in both the longitudinal and survival models. These estimators require the selection of smoothing parameters. 

In the longitudinal model,
 smoothing parameters are selected via restricted Maximum Likelihood (REML) at each iteration of the EM algorithm. By reformulating the estimation of smooth functions as  weighted least squares problems, computation can be efficiently carried out using the \textit{gam} function in the R package \textit{mgcv} \citep{Wood2011,Wood2016} (see Web Appendix D for details). 
 
In the survival model, the smoothing parameter associated with the coefficient function is selected locally at each iteration of the EM algorithm using the Akaike information criterion (AIC), $$\textnormal{AIC}=-2 \sum^n_{i=1}\mathbb{E}_i \left\{\log\big(f(T_i,\Delta_i|\widehat{h}_0, \boldsymbol{Z}_i,W_i(s),\boldsymbol{\xi}_i, \widehat{\boldsymbol{\gamma}}_{2},\widehat{\bomega}_2,\widehat{\boldsymbol{\gamma}}_{3})\big)\right\}+ 2\cdot \mathrm{tr}(\boldsymbol{I}^{-1}_{\widehat{\boldsymbol{\gamma}}}\boldsymbol{I}_{\widehat{\bgamma},0}),$$
where 
%$\sum^n_{i=1}\mathbb{E}_i \{\log\big(f(T_i,\Delta_i|\widehat{h}_0, \boldsymbol{Z}_i,W_i(s),\boldsymbol{\xi}_i, \widehat{\boldsymbol{\gamma}}_{2},\widehat{\bomega}_2,\widehat{\boldsymbol{\gamma}}_{3})\big)\}$ is the expected survival log-likelihood evaluated at the current estimates, 
$\mathrm{tr}(\boldsymbol{I}^{-1}_{\widehat{\boldsymbol{\gamma}}}\boldsymbol{I}_{\widehat{\bgamma},0})$ represents the effective degrees of freedom of  $\bgamma$, and $\boldsymbol{I}_{\bgamma,0}$ is the information matrix without the smoothness penalty term. 
The smoothing parameter $\tau_2^{\beta}$ is selected from 
a grid $\{e^{-10},e^{-9},\ldots,e^{0},\ldots,e^{9},e^{10}\}$, scaled by  the mean of the diagonal elements of $\boldsymbol{I}_{\bgamma,0}$ evaluated at the previous iteration. Although smoothing parameters are updated iteratively, they stabilize rapidly as the EM algorithm converges.

The number of functional principal components $L$ is selected using the Bayesian information criteria (BIC), $$\textnormal{BIC}=-2 \ell(\widehat{\boldsymbol{\Omega}} )+\log (n)\cdot \mathrm{df},$$
where
$\ell(\widehat{\boldsymbol{\Omega}})$ is the marginal log-likelihood, approximated by
\begin{equation*}
\sum_{i=1}^n \Bigg[\log\big\{f(\boldsymbol{y}_i|\boldsymbol{Z}_i, W_i(s),\widehat{\boldsymbol{\Omega}})\big\} + \log\Big\{ R^{-1} \sum_{r=1}^R f\left(T_i, \Delta_i \mid \widehat{h}_0, \boldsymbol{Z}_i, W_i(s),\boldsymbol{\xi}_i^{(r)}, \widehat{\boldsymbol{\gamma}}_2, \widehat{\bomega}_2,\widehat{\boldsymbol{\gamma}}_3 \right)\Big\} \Bigg] \;.
\end{equation*}
Here, $f(\boldsymbol{y}_i|\boldsymbol{Z}_i, W_i(s),\widehat{\boldsymbol{\Omega}})$ is the marginal density of the longitudinal data (multivariate normal; see Web Appendix B), and $\boldsymbol{\xi}_i^{(r)}$ are samples  from $f(\boldsymbol{\xi}_i|\boldsymbol{y}_i, \boldsymbol{Z}_i, W_i(s),\boldsymbol{\widehat{\Omega}})$ (Web Appendix C). 
A larger Monte Carlo sample size 
$R$
R is used than in the EM algorithm to ensure accurate approximation.

The degrees of freedom (DOF) are computed as
\begin{equation*}
 \mathrm{df}_{\mu} + \mathrm{df}_{\beta_1} + \mathrm{df}_{\beta_2} + \left\{ \sum_{\ell=1}^{L} \mathrm{df}_{\phi_{\ell}} - \frac{L(L-1)}{2}\right\}  + 2P + 1 + L,
\end{equation*}
where the terms correspond to the mean function, coefficient functions in the longitudinal and survival models, eigenfunctions of the covariance, regression coefficients and error variance, and the FPC effects in the survival model, respectively.

%\begin{itemize}
%    \item $\mathrm{df}_{\mu}$: DOF for estimating the mean function in the longitudinal model.
    
 %   \item $\mathrm{df}_{\beta_1} + \mathrm{df}_{\beta_2}$: DOF for estimating the coefficient function for the cumulative sitting profile $W_i(s)$ in the longitudinal model and Cox regression. 

  %  \item $ \sum_{\ell=1}^{L} \mathrm{df}_{\phi_{\ell}} - \frac{L(L-1)}{2}$: DOF for the covariance function's eigenpairs.  

   % \item $2P+1$: DOF for estimating coefficients for the baseline scalar covariates $\boldsymbol{Z}_i$ in the  longitudinal model and Cox regression and error variance in the longitudinal model. 

    %\item $L$: Number of coefficients for FPCA scores in the Cox regression.
%\end{itemize}
The optimal value of 
$L$ is selected over a grid by balancing model fit and complexity.

\section{Application to the OPACH Study} \label{sec:application}
We apply the proposed functional joint model (FJM) to the OPACH data, including 5,706 women with complete data. The longitudinal outcome is the SF-36 physical function score,  with higher values indicating better function, and the event outcome is all-cause mortality. Baseline scalar covariates include race (White, Black, and Hispanic), age, and awake time (hours being awake); see Table S1 in Web Appendix E for  details. The baseline functional covariate is the daily sitting bout accumulation profile. Longitudinal time is measured in years since enrollment in the OPACH study. All covariates, including the functional covariate, are centered so that  $\mu(t)$ represents the mean function.
 
Using the BIC criteria described in Section~\ref{sec:model.selection}, the proposed FJM selected $L=3$ functional principal components for  the longitudinal process. 
Figure~\ref{fig:application:mean} displays the estimated  mean function, which shows a decreasing trend over time, reflecting the age-related decline in physical function.
Figure~\ref{fig:app:eigenfunctions}  presents the estimated eigenfunctions. The first eigenfunction captures overall between-subject variability and explains the majority of the variation.
 The second eigenfunction reflects differences in the rate of decline, distinguishing subjects with accelerating (negative scores) versus decelerating (positive scores) trajectories.
 The third eigenfunction accounts for only  2.4\% of the variation and is less  interpretable.
%Figure~\ref{fig:application:mean} shows the estimated mean trajectory of longitudinal outcomes. 

\begin{figure}[!h]
	\centering
	\begin{subfigure}{0.45\textwidth}
	\centering
	\includegraphics[width=\linewidth]{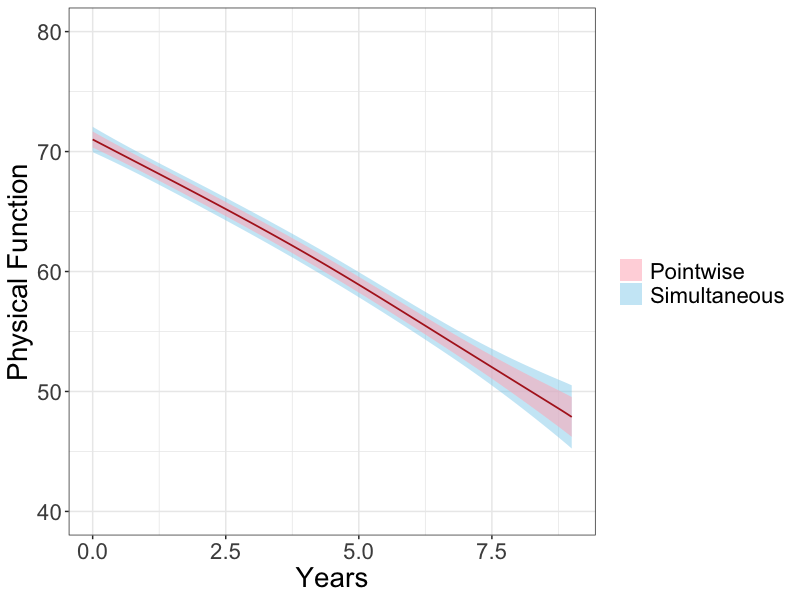}
	\caption{Mean function $\widehat{\mu}(t)$} 
	\label{fig:application:mean}
	\end{subfigure}%
	%\vskip 0.2\baselineskip
	\begin{subfigure}{0.45\textwidth}
	\centering
	\includegraphics[width=\linewidth]{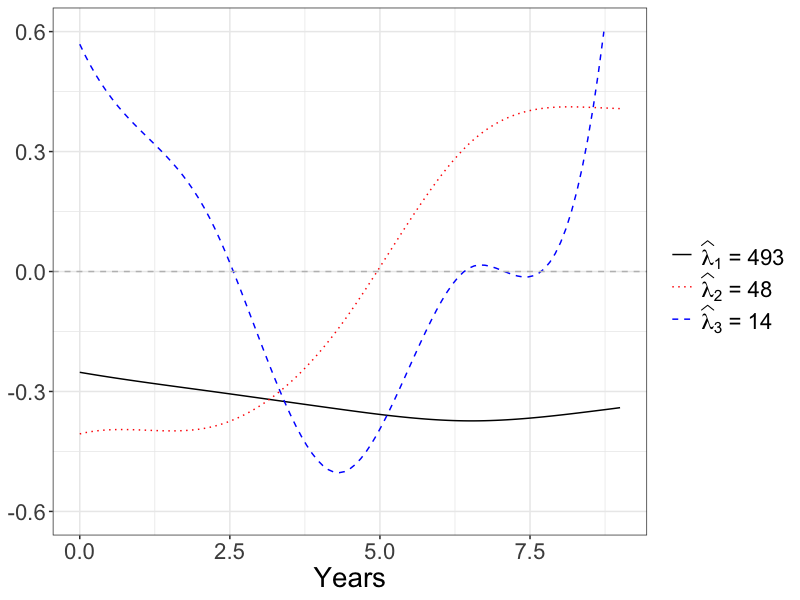}
	\caption{Eigenfunctions $\widehat{\phi}_\ell(t),\, \ell=1, 2, 3$} 
	    \label{fig:app:eigenfunctions}
	\end{subfigure}%
	\caption{Estimated mean function (red line, left panel) with 95\% pointwise (simultaneous) confidence bands and estimated eigenfunctions (right panel) with estimated eigenvalues for modeling SF-36 physical function score in the longitudinal model.}
	 %\label{fig:application:functional.covariate}
\end{figure}

Next, we examine the association between daily sitting bout accumulation profiles and longitudinal and survival outcomes. 
Figure~\ref{fig:application:functional.covariate} presents the estimated coefficient functions along with the  95\% pointwise confidence bands and the simultaneous confidence bands \citep{Ruppert_Wand_Carroll_2003}.
In the longitudinal model (Figure~\ref{fig:application:functional.covariate}a), the estimated coefficient function is negative across all bout durations, indicating that longer sitting time is associated with poorer physical function. Moreover, the magnitude of this association increases with bout duration, suggesting that prolonged uninterrupted sitting has a stronger detrimental effect. For example, a total of 100 minutes of sitting accumulated in a single bout is associated with approximately 15\% greater decline in physical function compared to the same total time accumulated in two 50-minute bouts. This duration-specific effect would not be captured by models based solely on total sitting time.

In the survival model (Figure~\ref{fig:application:functional.covariate}b), the estimated coefficient function $\widehat{\beta}_2(s)$
 is positive across all durations, indicating that longer sitting time is associated with increased mortality risk. The increasing trend further suggests that longer sitting bouts are associated with higher risk, even when total sitting time is held constant.

\begin{figure}[!h]
	\centering
	\begin{subfigure}{0.45\textwidth}
	\centering
	\includegraphics[width=\linewidth]{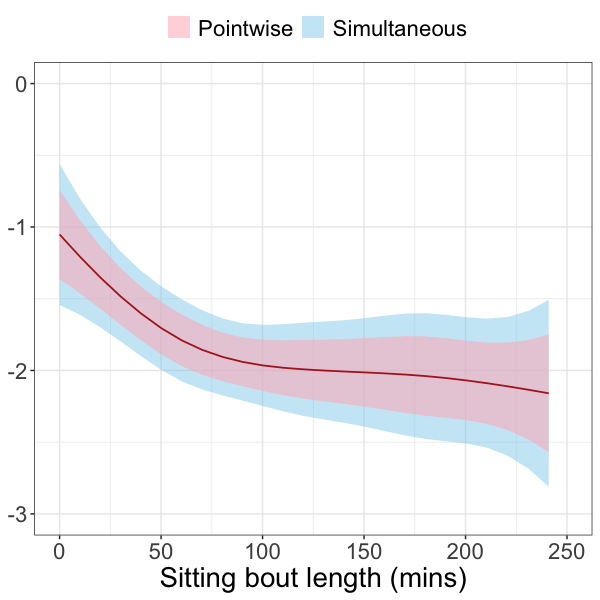}
	\caption{$\widehat{\beta}_1(s)$} 
	%\label{fig:nhanes_level1}
	\end{subfigure}%
	%\vskip 0.2\baselineskip
	\begin{subfigure}{0.45\textwidth}
	\centering
	\includegraphics[width=\linewidth]{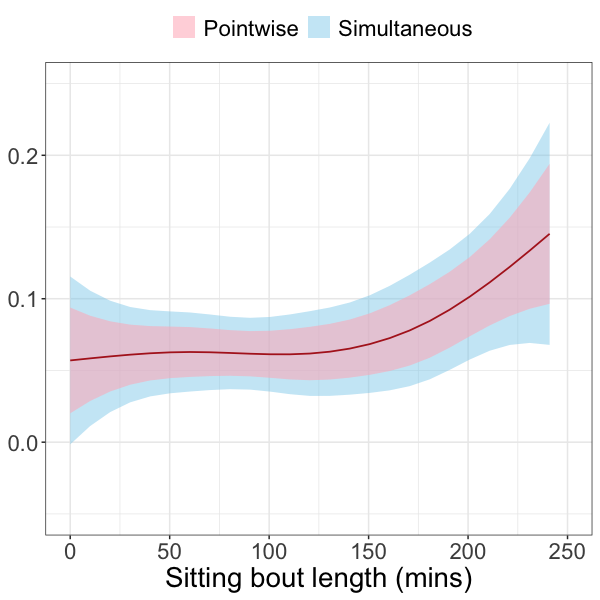}
	\caption{$\widehat{\beta}_2(s)$} 
	%\label{fig:nhanes_level2}
	\end{subfigure}%
	\caption{Estimated coefficient functions (red line) with 95\% point-wise confidence bands and  95\% simultaneous confidence bands in the  longitudinal model (left panel) and the survival model (right panel).}
	 \label{fig:application:functional.covariate}
\end{figure}

Table~\ref{table:app:scalar.covariate} reports the estimated effects of baseline scalar covariates and the FPC scores. The estimated covariate effects are consistent with established findings in aging research. The first two FPC scores are significantly associated with mortality and provide interpretable links between longitudinal trajectories and survival. Specifically, the first FPC score has a positive coefficient, while the corresponding eigenfunction is predominantly negative (Figure \ref{fig:app:eigenfunctions}), implying that individuals with lower scores exhibit better physical function and lower mortality risk. The second FPC score has a negative coefficient; combined with its eigenfunction, this indicates that individuals with negative scores experience accelerated functional decline and elevated mortality risk.

\begin{table}
\centering
\caption{Estimated parameters for baseline scalar covariates and SF-36 physical function scores in longitudinal model and estimated Cox regression coefficients (with standard errors).%, evaluating the association between baseline scalar covariates, FPCA scores, and time to event. 
Baseline covariates include race, age at baseline, and awake time. An asterisks indicates significance at the 0.05 level.}
\begin{tabular}{lllll}
        \hline
        \hline
         & \multicolumn{2}{c}{\text{Longitudinal}} &  \multicolumn{2}{c}{\text{Survival}}  \\
	\hline   
         Covariates &  \text{Estimate (s.e.)} & \text{P-value} & \text{Estimate (s.e.)} & \text{P-value} \\
	\hline 
         \text{Race: Hispanic}  & -0.089 (0.879)   & 0.919 & 0.373 (0.106)*&  0.000 \\
         \text{Race: Black}  &  -1.642 (0.866) & 0.058 &  0.217 (0.113) & 0.006  \\
         \text{Age}   & -1.106 (0.050)* & 0.000 &  0.104 (0.005)* & 0.000 \\
         \text{Awake time (hours per day)}   & 4.939 (0.263)* & 0.000 &  -0.228 (0.026)* & 0.000  \\
         \text{SF-36 physical function score ($\xi_{i1}$)} & - & - & 0.024 (0.001)* & 0.000  \\
         \text{SF-36 physical function score ($\xi_{i2}$)} & - & - & -0.019 (0.007)* & 0.006  \\
         \text{SF-36 physical function score ($\xi_{i3}$)} & - & - & 0.026 (0.016) & 0.102  \\
    %     \text{SF-36 physical function score ($\xi_{i4}$)} & - & - & 0.03 (0.03) & 0.24  \\
	\hline
\end{tabular}
\label{table:app:scalar.covariate}
\vskip18pt
\end{table}

\section{Simulation Study} \label{sec:simulation}

\subsection{Simulation settings}
We evaluate the performance of the proposed FJM model through simulations designed to mimic the OPACH data. Longitudinal data are generated  from  Model~\eqref{eq:longitudinal} using covariates from real data. For simplicity, we set the number of functional principal components to $L=2$. The mean function, coefficient functions, eigenfunctions, regression coefficients, and error variance are specified using estimates from the data application.

The  random scores $\xi_{i1}$ and $\xi_{i2}$ are generated independently from $\mathcal{N}(0,\lambda_{1\ell})$, where  $\lambda_{11}$ and  $\lambda_{12}$ are estimates from the real data. Measurement errors $\epsilon_{ij}$ are generated as  $\mathcal{N}(0,\sigma^2)$, with $\sigma^2$ taken from the real data analysis. Observation times $t_{ij}$ are set to 11 fixed equally spaced points on $[0,1]$ and are subject to right truncation by censoring and the event.

Event times are generated from the Cox regression model \eqref{eq:hazard}, incorporating random scores, scalar baseline covariates, and the functional  covariate.  The linear predictor is specified as $\boldsymbol{Z}_i^T\boldsymbol{\gamma}_{2} +\Big(\int_{s \in \mathcal{S}} \boldsymbol{b}^{\beta}(s)\mathrm{d} W_i(s) \Big) ^{\top}\bomega_2 + \xi_{i1} \gamma_{31} + \xi_{i2} \gamma_{32}$, with coefficients set to their estimated values from the data application.
%Specifically, $\boldsymbol{\gamma}_{11}=(0.02, 0.14, 0.05, 0.30)^T$, $\boldsymbol{\gamma}_{12}=(0.02, 0.13, 0.09, 0.35)^T$, and $\boldsymbol{\gamma}_{13}=(0.02, -0.19, 0.07, 0.40)^T$, with shared coefficients $\gamma_{211}=0.53$, $\gamma_{212}=-1.04$, $\gamma_{221}=0.45$, $\gamma_{222}=-1.11$, $\gamma_{231}=0.59$, and $\gamma_{232}=-0.09$. 
The baseline hazard is specified as Weibull, $h_0(t)= \rho t^{\rho-1}$, with $\rho=20$, and event times $S_i$ are generated using the inverse probability method \citep{bender2005generating}. Censoring times $C_i$ are independently sampled  from a Beta distribution, with parameters $\alpha_0$ and $\beta_0$ chosen to yield approximately 77\% censoring rates, matching the real data. For each subject, observations are retained only for $t_{ij} \leq T_i = \min(S_i, C_i)$ are retained. The sample size is 5708, the number of subjects in the real data,  with an average of $8.9$ observations per subject. We simulate data 100 times.

\subsection{Simulation results}
We first fit the FJM using the true number of components $L=2$. Figure~\ref{fig:simu:mean.eigen} displays the estimated mean functions and eigenfunctions across replications. %Figure~\ref{fig:simu:combined.eigen.vaule.function} presents the estimated eigenvalues and eigenfunctions across 100 replications. 
Figure~\ref{fig:simu:coefficient.function} presents the estimated coefficient functions in the longitudinal and  survival models. Gray curves corresponds to estimates from individual replications, dashed blue curved denote their averages, and solid red curves are the true functions.
The average estimates closely match the true functions, with only minor bias observed for $\widehat{\beta}_1(t)$.
A detailed summary of parameter estimates is provided in Web Appendix F, including eigenvalues, regression coefficients in both sub-models, and the error variance. Across all parameters, the estimates are close to the true values, indicating good finite-sample performance of the proposed method.

We further assess model selection for the number of principal components using AIC and BIC. The correct selection rates are 0.93 for AIC and 1.00 for BIC, demonstrating strong performance, with BIC showing superior reliability.

Overall, the simulation results confirm that the proposed FJM provides accurate parameter estimation and reliable model selection under realistic data-generating settings.

%\begin{figure}[h]
%    \centering
%    \includegraphics[width=0.5\textwidth]{figures/simu_mean_function.png}
%    \caption{Estimated mean functions from 100 simulation replications. Gray lines: individual estimates; dashed blue lines: the average of these estimates; solid red lines: true mean functions.}
%    \label{fig:simu:mean}
%\end{figure}

%\begin{figure}[h]
%    \centering
%    \includegraphics[width=1\textwidth]{figures/simu_mean_eigen.png}
%    \caption{Estimated mean functions and eigenfunctions from 100 simulation replications. Gray lines: estimates from 100 replicates; dashed blue lines: the average of these estimates; solid red lines: true mean function and eigenfunctions.}
%    \label{fig:simu:mean.eigen}
%\end{figure}

\begin{figure}[!h]
	\centering
	\begin{subfigure}{0.33\textwidth}
	\centering
	\includegraphics[width=\linewidth]{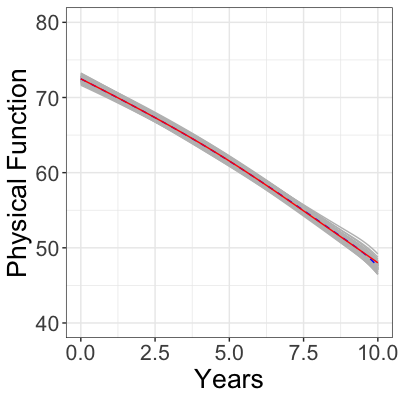}
	\caption{Mean function $\widehat{\mu}(t)$} 
	%\label{fig:application:mean}
	\end{subfigure}%
	%\vskip 0.2\baselineskip
	\begin{subfigure}{0.33\textwidth}
	\centering
	\includegraphics[width=\linewidth]{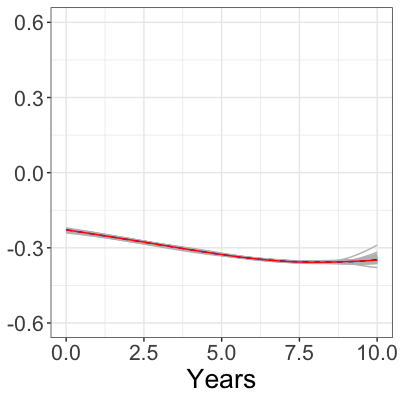}
	\caption{Eigenfunction $\widehat{\phi}_1(t)$} 
	    %\label{fig:app:eigenfunctions}
	\end{subfigure}%
        \begin{subfigure}{0.33\textwidth}
	\centering
	\includegraphics[width=\linewidth]{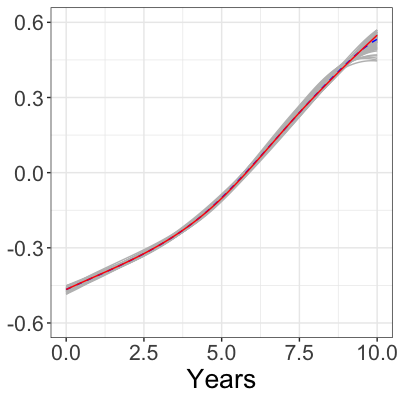}
	\caption{Eigenfunction $\widehat{\phi}_2(t)$} 
	    %\label{fig:app:eigenfunctions}
	\end{subfigure}%
	\caption{Estimated mean functions and eigenfunctions from 100 simulation replications. Gray curves: estimates from 100 replicates; dashed blue curves: average of these estimates; solid red curves: true mean function and eigenfunctions.}
	 \label{fig:simu:mean.eigen}
\end{figure}

%\begin{figure}[h]
%    \centering
%    % First panel (a)
%    \begin{subfigure}[t]{\textwidth}
%        \centering
%        \includegraphics[width=0.75\textwidth]{figures/simu_eigen_value.png}
%        \caption{Boxplots of estimated eigenvalues across 100 replicates.}
%        \label{fig:simu:eigenvalue}
%    \end{subfigure}
%    % Second panel (b)
%    \vspace{0.5cm}
%    \begin{subfigure}[t]{\textwidth}
%        \centering
%        \includegraphics[width=0.75\textwidth]{figures/simu_eigenfunction.png}
%        \caption{Estimated eigenfunctions across 100 replications. Gray lines: estimates from 100 replicates; dashed blue lines: the average estimated eigenfunctions; and solid red lines: the true eigenfunctions.}
%        \label{fig:simu:eigenfunction}
%    \end{subfigure}
%    % Overall caption for the figure
%    \caption{Plots of estimated eigenvalues and eigenfunctions across 100 replications.}
%    \label{fig:simu:combined.eigen.vaule.function}
%\end{figure}

%\begin{figure}[h]
%    \centering
%    \includegraphics[width=1\textwidth]{figures/simu_coefficient_function.png}
%    \caption{Estimated coefficient functions from 100 simulation replications. Gray lines: individual estimates; dashed blue lines: the average of these estimates; solid red lines: true coefficient functions.}
%    \label{fig:simu:coefficient.function}
%\end{figure}

\begin{figure}[!h]
	\centering
	\begin{subfigure}{0.45\textwidth}
	\centering
	\includegraphics[width=\linewidth]{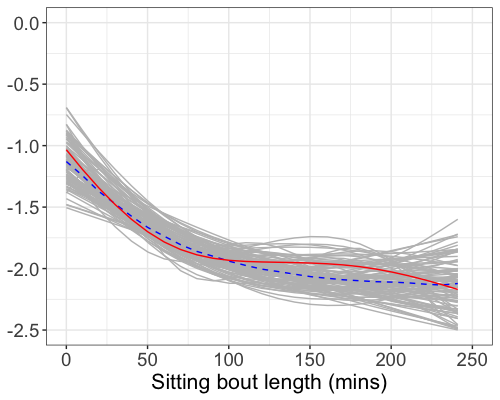}
	\caption{$\widehat{\beta}_1(s)$} 
	%\label{fig:nhanes_level1}
	\end{subfigure}%
	%\vskip 0.2\baselineskip
	\begin{subfigure}{0.45\textwidth}
	\centering
	\includegraphics[width=\linewidth]{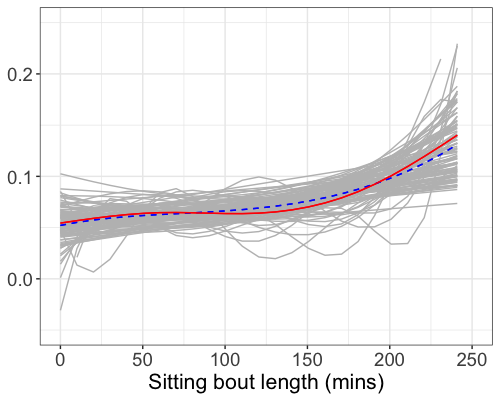}
	\caption{$\widehat{\beta}_2(s)$} 
	%\label{fig:nhanes_level2}
	\end{subfigure}%
	\caption{Estimated coefficient functions in the  longitudinal (left panel) and  survival models (right panel) from 100 simulation replications. Gray curves: individual estimates; dashed blue lines: average of these estimates; solid red lines: true coefficient functions.}
	 \label{fig:simu:coefficient.function}
\end{figure}

\section{Discussion} \label{sec:discussion}

We proposed a functional joint model (FJM) that incorporates a baseline functional covariate into both the longitudinal and survival sub-models. The model is motivated by the OPACH study and aims to characterize the relationship between daily sitting bout accumulation profiles, physical function trajectories, and all-cause mortality. By treating the entire sitting profile as a functional predictor, the proposed approach extends beyond traditional summary measures and enables a more detailed characterization of sedentary behavior and its association with health outcomes. 
%In the OPACH application, the model identifies interpretable, duration-specific associations with both longitudinal physical function and survival outcomes.

In the OPACH application, the model identifies interpretable, duration-specific associations with both physical function and survival. In both sub-models the estimated coefficient functions are non-constant, and their magnitude increases with bout duration. For a fixed amount of total daily sitting time, accumulating that time in longer, uninterrupted bouts is associated with worse physical function and greater mortality risk than accumulating the same total in shorter bouts. Such duration-specific patterns cannot be recovered from analyses based on simply summary metrics. These results add resolution to a growing body of evidence that prolonged, uninterrupted sedentary time is detrimental to health, and they align with public-health guidance to regularly interrupt sitting rather than to target total sitting time alone. More broadly, they illustrate the value of retaining the full distribution of bout durations as a functional covariate, as it allows estimation of flexible characterization of how the health effect is modified by bout length.

For estimation, we developed an EM-based algorithm that leverages penalized spline approximations for nonparametric components, including the coefficient functions associated with the functional covariate. A key feature of the proposed approach is the local selection of smoothing parameters within each EM iteration, which improves computational stability and flexibility. Compared with regression splines, penalized splines provide more stable estimation and yield smoother, more interpretable results.

This work contributes to the growing literature on functional joint modeling by addressing the incorporation of baseline functional covariates. Several limitations suggest directions for future research. First, the association between the longitudinal and survival processes is modeled through functional principal component scores, which imposes a parametric structure that may not capture more complex dependencies. Extensions to more flexible association structures, such as nonparametric formulations (e.g., \cite{zou2023bayesian}), may improve model performance. Second, the current framework considers a single baseline functional covariate. Extending the model to accommodate multiple or time-varying functional covariates would broaden its applicability in longitudinal and survival studies.

\clearpage\pagebreak\newpage
\baselineskip=14pt
\bibliographystyle{chicago}
\bibliography{ref}

@article{Dempster1977,
author = {Dempster, A. P. and Laird, N. M. and Rubin, D. B.},
title = {Maximum Likelihood from Incomplete Data Via the EM Algorithm},
journal = {Journal of the Royal Statistical Society: Series B (Methodological)},
volume = {39},
number = {1},
pages = {1-22},
year = {1977}
}

@article{li2022joint,
  title={Joint model for survival and multivariate sparse functional data with application to a study of Alzheimer's Disease},
  author={Li, Cai and Xiao, Luo and Luo, Sheng},
  journal={Biometrics},
  volume={78},
  number={2},
  pages={435--447},
  year={2022},
  publisher={Wiley Online Library}
}

@article{Eilers1996,
 author = {Paul H. C. Eilers and Brian D. Marx},
 journal = {Statistical Science},
 number = {2},
 pages = {89-102},
 publisher = {Institute of Mathematical Statistics},
 title = {Flexible Smoothing with $B$-splines and Penalties},
 volume = {11},
 year = {1996}
}

@article{Wood2011,
  title = {Fast stable restricted maximum likelihood and marginal likelihood estimation of semiparametric generalized linear models},
  author  = {Wood, S N},
  journal = {Journal of the Royal Statistical Society (B)},
  volume = {73},
  number =  {1},
  pages  = {3-36},
  year = {2011}
}

@article{Wood2016,
author = {Simon N. Wood and Natalya Pya and Benjamin Säfken},
title = {Smoothing Parameter and Model Selection for General Smooth Models},
journal = {Journal of the American Statistical Association},
volume = {111},
number = {516},
pages = {1548-1563},
year  = {2016},
publisher = {Taylor & Francis}
}

@article{bender2005generating,
  title={Generating survival times to simulate Cox proportional hazards models},
  author={Bender, Ralf and Augustin, Thomas and Blettner, Maria},
  journal={Statistics in medicine},
  volume={24},
  number={11},
  pages={1713--1723},
  year={2005},
  publisher={Wiley Online Library}
}

@article{wulfsohn1997joint,
  title={A joint model for survival and longitudinal data measured with error},
  author={Wulfsohn, Michael S and Tsiatis, Anastasios A},
  journal={Biometrics},
  pages={330--339},
  year={1997},
  publisher={JSTOR}
}

@article{yao2007functional,
  title={Functional principal component analysis for longitudinal and survival data},
  author={Yao, Fang},
  journal={Statistica Sinica},
  pages={965--983},
  year={2007},
  publisher={JSTOR}
}

@article{yan2017dynamic,
  title={Dynamic prediction of disease progression for leukemia patients by functional principal component analysis of longitudinal expression levels of an oncogene},
  author={Yan, Fangrong and Lin, Xiao and Huang, Xuelin},
  year={2017}
}

@article{zou2023bayesian,
  title={Bayesian inference and dynamic prediction for multivariate longitudinal and survival data},
  author={Zou, Haotian and Zeng, Donglin and Xiao, Luo and Luo, Sheng},
  journal={The Annals of Applied Statistics},
  volume={17},
  number={3},
  pages={2574--2595},
  year={2023},
  publisher={Institute of Mathematical Statistics}
}

@article{zou2024aoas,
  title={DYNAMIC PREDICTION WITH MULTIVARIATE LONGITUDINAL
OUTCOMES AND LONGITUDINAL MAGNETIC RESONANCE IMAGING
DATA},
  author={Zou, Haotian and Xiao, Luo and Zeng, Donglin  and Luo, Sheng},
  journal={The Annals of Applied Statistics},
  volume={},
  number={},
  pages={},
  year={2024},
  publisher={Institute of Mathematical Statistics},
note ={accepted}
}

@article{he2024joint,
  title={Joint mixed membership modeling of multivariate longitudinal and survival data for learning the individualized disease progression},
  author={He, Yuyang and Song, Xinyuan and Kang, Kai},
  journal={The Annals of Applied Statistics},
  volume={18},
  number={3},
  pages={1924--1946},
  year={2024},
  publisher={Institute of Mathematical Statistics}
}

@article{shi2024dynamic,
  title={Dynamic Survival Prediction using Sparse Longitudinal Images via Multi-dimensional Functional Principal Component Analysis},
  author={Shi, Haolun and Jiang, Shu and Ma, Da and Beg, Mirza Faisal and Cao, Jiguo},
  journal={Journal of Computational and Graphical Statistics},
  number={just-accepted},
  pages={1--18},
  year={2024},
  publisher={Taylor \& Francis}
}

@article{hong2021dynamic,
  title={Dynamic prediction of disease processes based on recurrent history and functional principal component analysis of longitudinal biomarkers: Application for ovarian epithelial cancer},
  author={Hong, Yizhou and Su, Liwen and Song, Siyi and Yan, Fangrong},
  journal={Statistics in Medicine},
  volume={40},
  number={8},
  pages={2006--2023},
  year={2021},
  publisher={Wiley Online Library}
}

@article{jiang2024functional,
  title={Functional partial least squares with censored outcomes: Prediction of breast cancer risk with mammogram images},
  author={Jiang, Shu and Cao, Jiguo and Colditz, Graham A},
  journal={The Annals of Applied Statistics},
  volume={18},
  number={2},
  pages={1051--1063},
  year={2024},
  publisher={Institute of Mathematical Statistics}
}

@article{jiang2023predicting,
  title={Predicting the onset of breast cancer using mammogram imaging data with irregular boundary},
  author={Jiang, Shu and Cao, Jiguo and Colditz, Graham A and Rosner, Bernard},
  journal={Biostatistics},
  volume={24},
  number={2},
  pages={358--371},
  year={2023},
  publisher={Oxford University Press}
}

@article{gellar2015cox,
  title={Cox regression models with functional covariates for survival data},
  author={Gellar, Jonathan E and Colantuoni, Elizabeth and Needham, Dale M and Crainiceanu, Ciprian M},
  journal={Statistical modelling},
  volume={15},
  number={3},
  pages={256--278},
  year={2015},
  publisher={SAGE Publications Sage India: New Delhi, India}
}

@article{kong2018flcrm,
  title={FLCRM: Functional linear cox regression model},
  author={Kong, Dehan and Ibrahim, Joseph G and Lee, Eunjee and Zhu, Hongtu},
  journal={Biometrics},
  volume={74},
  number={1},
  pages={109--117},
  year={2018},
  publisher={Wiley Online Library}
}

@article{wang2020partial,
  title={Partial least squares for functional joint models with applications to the Alzheimer's disease neuroimaging initiative study},
  author={Wang, Yue and Ibrahim, Joseph G and Zhu, Hongtu},
  journal={Biometrics},
  volume={76},
  number={4},
  pages={1109--1119},
  year={2020},
  publisher={Wiley Online Library}
}

@article{cui2021additive,
  title={Additive functional Cox model},
  author={Cui, Erjia and Crainiceanu, Ciprian M and Leroux, Andrew},
  journal={Journal of Computational and Graphical Statistics},
  volume={30},
  number={3},
  pages={780--793},
  year={2021},
  publisher={Taylor \& Francis}
}

@article{qu2016optimal,
  title={Optimal estimation for the functional cox model},
  author={Qu, Simeng and Wang, Jane-Ling and Wang, Xiao},
  year={2016}
}

@book{Ruppert_Wand_Carroll_2003, place={Cambridge}, series={Cambridge Series in Statistical and Probabilistic Mathematics}, title={Semiparametric Regression}, publisher={Cambridge University Press}, author={Ruppert, David and Wand, M. P. and Carroll, R. J.}, year={2003}, collection={Cambridge Series in Statistical and Probabilistic Mathematics}}

@article{tsiatis2004joint,
  title={Joint modeling of longitudinal and time-to-event data: an overview},
  author={Tsiatis, Anastasios A and Davidian, Marie},
  journal={Statistica Sinica},
  pages={809--834},
  year={2004},
  publisher={JSTOR}
}

@article{yang2010review,
  title={A review of accelerometry-based wearable motion detectors for physical activity monitoring},
  author={Yang, Che-Chang and Hsu, Yeh-Liang},
  journal={Sensors},
  volume={10},
  number={8},
  pages={7772--7788},
  year={2010},
  publisher={Molecular Diversity Preservation International (MDPI)}
}

@article{greenwood2021cnn,
  title={The CNN Hip Accelerometer Posture (CHAP) method for classifying sitting patterns from hip accelerometers: A validation study},
  author={Greenwood-Hickman, Mikael Anne and Nakandala, Supun and Jankowska, Marta M and Rosenberg, Dori E and Tuz-Zahra, Fatima and Bellettiere, John and Carlson, Jordan and Hibbing, Paul R and Zou, Jingjing and Lacroix, Andrea Z and others},
  journal={Medicine and science in sports and exercise},
  volume={53},
  number={11},
  pages={2445},
  year={2021}
}

@article{eanes2018too,
  title={CE: Too much sitting: A newly recognized health risk},
  author={Eanes, Linda},
  journal={AJN The American journal of nursing},
  volume={118},
  number={9},
  pages={26--34},
  year={2018},
  publisher={LWW}
}

@article{women1998design,
  title={Design of the Women’s Health Initiative clinical trial and observational study},
  author={Women's Health Initiative Study Group and others},
  journal={Controlled clinical trials},
  volume={19},
  number={1},
  pages={61--109},
  year={1998},
  publisher={Elsevier BV}
}

@article{lacroix2017objective,
  title={The objective physical activity and cardiovascular disease health in older women (OPACH) study},
  author={LaCroix, Andrea Z and Rillamas-Sun, Eileen and Buchner, David and Evenson, Kelly R and Di, Chongzhi and Lee, I-Min and Marshall, Steve and LaMonte, Michael J and Hunt, Julie and Tinker, Lesley Fels and others},
  journal={BMC public health},
  volume={17},
  pages={1--12},
  year={2017},
  publisher={Springer}
}

@article{wang2025multi,
author = {Wang, Wenyi and Xiao, Luo and Li, Ruonan and Luo, Sheng and Alzheimer's Disease Neuroimaging Initiative},
title = {A Functional Joint Model for Survival and Multivariate Sparse Functional Data in Multi-Cohort Alzheimer's Disease Study},
journal = {Statistics in Medicine},
volume = {45},
number = {3-5},
pages = {e70442},
doi = {https://doi.org/10.1002/sim.70442},
url = {https://onlinelibrary.wiley.com/doi/abs/10.1002/sim.70442},
eprint = {https://onlinelibrary.wiley.com/doi/pdf/10.1002/sim.70442},
year = {2026}
}

@article{Louis1982,
    author = {Louis, Thomas A.},
    title = {Finding the Observed Information Matrix When Using the EM Algorithm},
    journal = {Journal of the Royal Statistical Society: Series B (Methodological)},
    volume = {44},
    number = {2},
    pages = {226-233},
    year = {2018},
    month = {12},
    issn = {0035-9246},
    doi = {10.1111/j.2517-6161.1982.tb01203.x},
    url = {https://doi.org/10.1111/j.2517-6161.1982.tb01203.x},
    eprint = {https://academic.oup.com/jrsssb/article-pdf/44/2/226/49097595/jrsssb_44_2_226.pdf},
}

\clearpage\pagebreak\newpage
\pagestyle{fancy}
\fancyhf{}
\rhead{\bfseries\thepage}
\lhead{\bfseries NOT FOR PUBLICATION SUPPLEMENTARY MATERIAL}
\begin{center}
%{\LARGE{\bf Supplementary Material to\\  {\it Fast Multilevel Functional Principal Component Analysis}}}
{\LARGE{\bf Supplementary Material to\\  {\it A functional joint model with baseline functional covariates: linking sitting accumulation patterns to physical function and mortality among older women}}}
\end{center}

\baselineskip=12pt

\vskip 1cm
\begin{center}

Luo Xiao\\
Department of Statistics, North Carolina State University, Raleigh, North Carolina\\
\hskip 5mm \\

Wenyi Wang\\
Department of Statistics, North Carolina State University, Raleigh, North Carolina\\
\hskip 5mm \\

Yumeng Zhang\\
Department of Statistics, North Carolina State University, Raleigh, North Carolina\\
\hskip 5mm \\

%Ilsuk Kang \\
%Public Health Sciences Division, Fred Hutch Cancer Center, Seattle, Washington\\
%\hskip 5mm \\

Mike Lamonte\\
Department of Epidemiology and Environmental Health, University of Buffalo, Buffalo, NY\\
\hskip 5mm

Andrea LaCroix\\
Herbert Wertheim School of Public Health \& Human Longevity Science, University of California, San Diego, California\\
\hskip 5mm

Chongzhi Di\\
Public Health Sciences Division, Fred Hutchinson Cancer Center, Seattle, Washington\\

\end{center}

\setcounter{figure}{0}
\setcounter{equation}{0}
\setcounter{page}{1}
\setcounter{table}{0}
\setcounter{section}{0}
\renewcommand{\thefigure}{S\arabic{figure}}
\renewcommand{\theequation}{S\arabic{equation}}
\renewcommand{\thesection}{S.\arabic{section}}
\renewcommand{\thesubsection}{S.\arabic{section}.\arabic{subsection}}
\renewcommand{\thepage}{S.\arabic{page}}
\renewcommand{\thetable}{S\arabic{table}}
\baselineskip=17pt

\newpage

\section*{Web Appendix A: Sitting Profile Formulation}\label{sec:point}

Suppose that subject $i$ has $m_i$ days of accelerometry data.
For day $j$, let $\{s^{[j]}_{ik}, k =1,\ldots, K_i^{[j]}\}$ be the collection of sitting bout durations with a total of  $K_i^{[j]}$ sitting bouts. Then
$N_i^{[j]}(s)= \sum_{k=1}^{K_i^{[j]}} \mathbb{I}(s_{ik}^{[j]}\leq s)$
and $N_i(s) = m_i^{-1} \sum_{j=1}^{m_i}  \mathrm{d} N_i^{[j]}(s)$. Finally, the cumulative profile $W_i(s)$ is formulated so that $\mathrm{d} W_i(s) = s \,\mathrm{d} N_i(s)$.

\section*{Web Appendix B: Joint Distribution of Longitudinal Outcomes and FPC Scores}\label{app:joint}

We derive the joint distribution of
$f(\boldsymbol{y}_{i},\boldsymbol{\xi}_{i}|\boldsymbol{\Lambda})$, which characterizes the relationship between the observed longitudinal outcomes and the functional principal component scores. The observed outcomes follow the equation: $\boldsymbol{y}_{i}=\boldsymbol{\mu}_{i} + \boldsymbol{F}_i + \boldsymbol{\Phi}_{i}\boldsymbol{\xi}_{i} + \boldsymbol{\epsilon}_{i}$, where $\boldsymbol{\epsilon}_{i} = (\epsilon_{i1},\ldots,\epsilon_{im_{i}})^{\top}$ is the vector of measurement errors. We have
\begin{equation*}
\begin{pmatrix} \boldsymbol{y}_i \\ \boldsymbol{\xi}_i \end{pmatrix} \sim \mathcal{N}\left[
\begin{pmatrix} \boldsymbol{\mu}_{i}+\boldsymbol{F}_i\\\boldsymbol{0}\end{pmatrix},\;
\begin{pmatrix} \text{Cov}(\boldsymbol{y}_i) & \text{Cov}(\boldsymbol{y}_i,\boldsymbol{\xi}_i) \\ \text{Cov}(\boldsymbol{\xi}_i,\boldsymbol{y}_i) &\text{Cov}(\boldsymbol{\xi}_i)\end{pmatrix}
\right] \;,
\end{equation*}
where $\text{Cov}(\boldsymbol{y}_i)=\boldsymbol{\Phi}_{i}\boldsymbol{\Lambda}\boldsymbol{\Phi}^{\top}_{i}+\sigma^2\boldsymbol{I}_{m_{i}}=\boldsymbol{\Phi}_{i}\boldsymbol{\Lambda}\boldsymbol{\Phi}^{\top}_{i}+\boldsymbol{\Sigma}_i$, $\text{Cov}(\boldsymbol{y}_i,\boldsymbol{\xi}_i)=\boldsymbol{\Phi}_{i}\boldsymbol{\Lambda}$ and $\text{Cov}(\boldsymbol{\xi}_i)=\bLambda$. Using conditional expectation, the posterior mean and covariance of $\boldsymbol{\xi}_i$ given $\boldsymbol{y}_i$ are derived as: $\mathbb{E}_i(\boldsymbol{\xi}_i|\boldsymbol{y}_i)=\text{Cov}(\boldsymbol{\xi}_i,\boldsymbol{y}_i)\text{Cov}(\boldsymbol{y}_i)^{-1}(\boldsymbol{y}_i-\boldsymbol{\mu}_i-\boldsymbol{F}_i)$, $\text{Cov}(\boldsymbol{\xi}_i|\boldsymbol{y}_i)=\text{Cov}(\boldsymbol{\xi}_i)-\text{Cov}(\boldsymbol{\xi}_i,\boldsymbol{y}_i)\text{Cov}^{-1}(\boldsymbol{y}_i)\text{Cov}(\boldsymbol{y}_i,\boldsymbol{\xi}_i)$.

\section*{Web Appendix C: Marginal Log-likelihood} \label{app:marginal}
The marginal log-likelihood is a key quantity for model evaluation and comparison, providing a measure of how well the proposed model fits the observed data. Since the exact computation of this likelihood involves intractable integrals, we employ an approximation strategy as detailed below.

The marginal log-likelihood of the data is given by
\begin{equation*}
\begin{split}
&\ell(\widehat{\boldsymbol{\Omega}})  =\sum_{i=1}^n \log \left\{\int f\left(\boldsymbol{y}_i \mid \boldsymbol{x}_i, \widehat{\boldsymbol{\Sigma}}_i\right) f\left(T_i,\Delta_i \mid \widehat{h}_0, \boldsymbol{Z}_i,W_i(s),\boldsymbol{\xi}_i, \widehat{\boldsymbol{\gamma}}_{2},\widehat{\bomega}_2,\widehat{\boldsymbol{\gamma}}_{3}\right) f\left(\boldsymbol{\xi}_{i} \mid \widehat{\boldsymbol{\Lambda}}\right) \mathrm{d} \boldsymbol{\xi}_i\right\} \\
& =\sum_{i=1}^n \log \left\{f\left(\boldsymbol{y}_i \mid \boldsymbol{Z}_i,W_i(s),\widehat{\boldsymbol{\Omega}}\right) \int f\left(\boldsymbol{\xi}_i \mid \boldsymbol{y}_i, \boldsymbol{Z}_i,W_i(s), \widehat{\boldsymbol{\Omega}}\right) f\left(T_i, \Delta_i \mid  \widehat{h}_0, \boldsymbol{Z}_i,W_i(s),\boldsymbol{\xi}_i, \widehat{\boldsymbol{\gamma}}_{2},\widehat{\bomega}_2,\widehat{\boldsymbol{\gamma}}_{3}\right) \mathrm{d} \boldsymbol{\xi}_i\right\}.    
\end{split}    
\end{equation*}

Since this integral does not have a closed-form solution, we approximate it using Monte Carlo integration. Specifically, the integral is approximated as: 
\begin{equation*}
\left\{R^{-1} \sum_{r=1}^R f\left(T_i, \Delta_i \mid \widehat{h}_0, \boldsymbol{Z}_i,W_i(s), \boldsymbol{\xi}_i^{(r)}, \widehat{\boldsymbol{\gamma}}_{2},\widehat{\bomega}_2,\widehat{\boldsymbol{\gamma}}_{3}\right)\right\},
\end{equation*} 
where $\boldsymbol{\xi}_i^{(r)}$ are samples drawn from the posterior distribution $f\left(\boldsymbol{\xi}_i \mid \boldsymbol{y}_i, \boldsymbol{Z}_i,W_i(s), \widehat{\boldsymbol{\Omega}}\right)$. Substituting this approximation, the marginal log-likelihood becomes:
%because there is no closed form for %$\ell(\widehat{\boldsymbol{\Omega}})$. Then, we have
\begin{equation*}
\ell(\widehat{\boldsymbol{\Omega}}) \approx \sum_{i=1}^n \left[\log \left\{f\left(\boldsymbol{y}_i \mid \boldsymbol{Z}_i, W_i(s),\widehat{\boldsymbol{\Omega}}\right)\right\}+\log \left\{R^{-1} \sum_{r=1}^R f\left(T_i, \Delta_i \mid \widehat{h}_0, \boldsymbol{Z}_i, W_i(s),\boldsymbol{\xi}_i^{(r)}, \widehat{\boldsymbol{\gamma}}_2, \widehat{\bomega}_2,\widehat{\boldsymbol{\gamma}}_3\right)\right\}\right].
\end{equation*}

\section*{Web Appendix D: Details on M-step for longitudinal model survival model}\label{app:Mstep}
We estimate $\boldsymbol{\alpha}$, $\bomega_1$ and $\boldsymbol{\Theta}$ using P-splines \citep{Eilers1996} with smoothing parameters. P-splines provide stable estimates that enhance the convergence speed of the iterative algorithm, contributing to computational efficiency. 

First, we estimate ${\boldsymbol{\alpha}}$, $\bgamma_1$,  and $\bomega_1$ by minimizing the following penalized least squares objective function:
\begin{equation}\label{eq:alpha}
\sum_{i=1}^n\Big\Vert\boldsymbol{y}_{i} -\widehat{\boldsymbol{\Phi}}_{i}\mathbb{E}_i(\boldsymbol{\xi}_{i})- \boldsymbol{B}_{i}\boldsymbol{\alpha}-\bZ_i^{\top}\bgamma_1\boldsymbol{1}_{m_i}- \Big(\int_{s\in\mathcal{S}} \boldsymbol{b}^{\beta}(s_{j_s})^{\top}\bomega_1\mathrm{d}W_i(s)\Big)\boldsymbol{1}_{m_i}\Big\Vert^2 + \tau_{\alpha}\boldsymbol{\alpha}^{\top}\boldsymbol{P} \boldsymbol{\alpha} + \tau_{1}^{\beta}\boldsymbol{\omega}_1^{\top}\boldsymbol{P}^{\beta} \boldsymbol{\omega}_1\;,
\end{equation}
where $\Vert\cdot\Vert$ is the Euclidean norm, $\boldsymbol{P}$ and $\boldsymbol{P}^{\beta}$ are two penalty matrices, $\tau_{\alpha}$ and $\tau_{1}^{\beta}$ are smoothing parameters. 

Second, given $\boldsymbol{\theta}_{1\ell'}$, $\ell'\neq \ell$, we estimate $\boldsymbol{\theta}_{\ell}$ by minimizing:
\begin{equation*}
    \sum_{i=1}^n\mathbb{E}_i \Big\Vert \boldsymbol{y}_{i}-\widehat{\boldsymbol{\mu}}_{i}-\bZ_i^{\top}\widehat{\bgamma}_1\boldsymbol{1}_{m_i}-\Big(\int_{s\in\mathcal{S}} \boldsymbol{b}^{\beta}(s_{j_s})^{\top}\bomega_1\mathrm{d}W_i(s)\Big)\boldsymbol{1}_{m_i} -\sum_{\ell'\neq \ell}\widetilde{\boldsymbol{B}}_{i}\widehat{\boldsymbol{\theta}}_{\ell'}{\xi}_{i \ell'} - \widetilde{\boldsymbol{B}}_{i}\boldsymbol{\theta}_{\ell}{\xi}_{i\ell}\Big\Vert^2 + \tau_{\theta,\ell}\boldsymbol{\theta}^{\top}_{\ell}\boldsymbol{P} \boldsymbol{\theta}_{\ell}\;.
\end{equation*}
This is equivalent to minimizing:
\begin{equation}\label{eq:theta1}
\begin{split}
\sum_{i=1}^n\Big\Vert &\frac{\mathbb{E}_i(\xi_{i \ell})}{\sqrt{\mathbb{E}_i(\xi^2_{i \ell})}}\Big(\boldsymbol{y}_{i}-\widehat{\boldsymbol{\mu}}_{i}-\bZ_i^{\top}\widehat{\bgamma}_1\boldsymbol{1}_{m_i}-\Big(\int_{s\in\mathcal{S}} \boldsymbol{b}^{\beta}(s_{j_s})^{\top}\bomega_1\mathrm{d}W_i(s)\Big)\boldsymbol{1}_{m_i}\Big) \\
&- \frac{1}{\sqrt{\mathbb{E}_i(\xi^2_{i \ell})}}\sum_{\ell'\neq \ell}\widetilde{\boldsymbol{B}}_{i}\widehat{\boldsymbol{\theta}}_{\ell'}\mathbb{E}_i(\xi_{i \ell'}{\xi}_{i \ell})  - \widetilde{\boldsymbol{B}}_{i}\boldsymbol{\theta}_{\ell}\sqrt{\mathbb{E}_i(\xi^2_{i \ell})}\Big\Vert^2 + \tau_{\theta,\ell}\boldsymbol{\theta}^{\top}_{\ell}\boldsymbol{P} \boldsymbol{\theta}_{\ell}\;. 
\end{split}
\end{equation}

Minimizing (\ref{eq:alpha}) and (\ref{eq:theta1}) can be solved directly through the {\it gam} function of the R package {\it mgcv} \citep{Wood2011,Wood2016}. The smoothing parameters are selected locally through Restricted Maximum Likelihood (REML) during each iteration. The procedure is repeated for each column of $\boldsymbol{\Theta}$ until convergence, ensuring a robust estimation.

Third, the variance of random noise $\sigma^2$ is updated as:
\begin{equation*} 
\begin{split}
    \widehat{\sigma}^2 &= \frac{1}{\sum_{i=1}^n m_{i}}\sum_{i=1}^n\mathbb{E}_i\Big\Vert \boldsymbol{y}_{i}-\widehat{\boldsymbol{\mu}}_{i}-\bZ_i^{\top}\widehat{\bgamma}_1\boldsymbol{1}_{m_i}-\Big(\int_{s\in\mathcal{S}} \boldsymbol{b}^{\beta}(s_{j_s})^{\top}\bomega_1\mathrm{d}W_i(s)\Big)\boldsymbol{1}_{m_i} - \widehat{\boldsymbol{\Phi}}_{i}\boldsymbol{\xi}_{i} \Big\Vert^2  \\
    & =  \frac{1}{\sum_{i=1}^n m_{i}}\sum_{i=1}^n \Bigg\{\Big\Vert\boldsymbol{y}_{i} - \widehat{\boldsymbol{\mu}}_{i}-\bZ_i^{\top}\widehat{\bgamma}_1\boldsymbol{1}_{m_i}-\Big(\int_{s\in\mathcal{S}} \boldsymbol{b}^{\beta}(s_{j_s})^{\top}\bomega_1\mathrm{d}W_i(s)\Big)\boldsymbol{1}_{m_i}\Big\Vert^2 -\\
    & 2\Big(\boldsymbol{y}_{i}-\widehat{\boldsymbol{\mu}}_{i}-\bZ_i^{\top}\widehat{\bgamma}_1\boldsymbol{1}_{m_i}-\Big(\int_{s\in\mathcal{S}} \boldsymbol{b}^{\beta}(s_{j_s})^{\top}\bomega_1\mathrm{d}W_i(s)\Big)\boldsymbol{1}_{m_i}\Big)^{\top}\big(\widehat{\boldsymbol{\Phi}}_{i}\mathbb{E}_i(\boldsymbol{\xi}_{i})\big) +\text{tr}\big\{\widehat{\boldsymbol{\Phi}}^{\top}_{i}\widehat{\boldsymbol{\Phi}}_{i}\mathbb{E}_i(\boldsymbol{\xi}_{i}\boldsymbol{\xi}^{\top}_{i})\big\} 
    \Bigg\} \;.
\end{split}
\end{equation*}

Finally, we derive the observed information matrix for $\boldsymbol{\theta}= (\boldsymbol{\alpha}^{\top}, \boldsymbol{\gamma}_1^{\top}, \bomega_1^{\top})^{\top}$ using Louis' formula \citep{Louis1982}.
Let the observed data for subject $i$ be $\boldsymbol{Y}_i=\{\boldsymbol{y}_i,\ \bZ_i,\ W_i(\cdot),\ T_i,\ \Delta_i\}$,
and let $\boldsymbol{Y}=\{\boldsymbol{Y}_i\}_{i=1}^n$ denote the observed data for the full sample.
%To conduct inference on $\boldsymbol{\theta}=(\boldsymbol{\alpha}^{\top},\bgamma_1^{\top},\bomega_1^{\top})^{\top}$ after EM convergence, we compute the observed information using the identity of \textcolor{red}{Louis1982}. 
Let $\ell_{1,c}$ denote the complete-data penalized longitudinal log-likelihood:
\begin{equation*}
\begin{split}
 \ell_{1,c}
&=-\sum_{i=1}^n\sum_{j=1}^{m_i}\Bigg \{ \frac{1}{2}\text{log}(2\pi \sigma^2)
+\frac{1}{2\sigma^2}\Big(y_{ij}-\boldsymbol{b}(t_{ij})^{\top}\boldsymbol{\alpha}-\bZ_i^{\top}\bgamma_1-\int_{s\in\mathcal{S}}\boldsymbol{b}^{\beta}(s)^{\top}\bomega_1\,\mathrm{d}W_i(s) \\
&\hspace{1cm}-\widetilde{\boldsymbol{b}}(t_{ij})^{\top}\boldsymbol{\Theta}\boldsymbol{\xi}_{i}\Big)^2
\Bigg\}
-\tau_{\alpha}\boldsymbol{\alpha}^{\top}\boldsymbol{P} \boldsymbol{\alpha}
-\tau_{1}^{\beta}\boldsymbol{\omega}_1^{\top}\boldsymbol{P}^{\beta}\boldsymbol{\omega}_1
-\sum_{\ell=1}^L\tau_{\theta,\ell}\boldsymbol{\theta}_{\ell}^{\top}\boldsymbol{P}\boldsymbol{\theta}_{\ell}.
\end{split}
\end{equation*}
Let $\boldsymbol{S}_{1,c}=\partial\ell_{1,c}/\partial\boldsymbol{\theta}$ denote the  score vector for the complete data.
Then
\begin{equation*}
\begin{split}
\boldsymbol{S}_{1,c}
&=
-\frac{1}{2\sigma^2}\sum_{i=1}^n\sum_{j=1}^{m_i}
\Bigg\{
2\Big(
y_{ij}-\boldsymbol{b}(t_{ij})^{\top}\boldsymbol{\alpha}
-\bZ_i^{\top}\bgamma_1
-\int_{s\in\mathcal{S}}\boldsymbol{b}^{\beta}(s)^{\top}\bomega_1\,\mathrm{d}W_i(s)
-\widetilde{\boldsymbol{b}}(t_{ij})^{\top}\boldsymbol{\Theta}\boldsymbol{\xi}_{i}
\Big) \\
&\hspace{0.5cm}
\Big(
-\boldsymbol{b}(t_{ij})^{\top},\ 
-\bZ_i^{\top},\ 
-\int_{s\in\mathcal{S}}\boldsymbol{b}^{\beta}(s)^{\top}\mathrm{d}W_i(s)
\Big)^{\top}
\Bigg\}
-
2\Big(
\tau_{\alpha}(\boldsymbol{P}\boldsymbol{\alpha})^{\top},\
\boldsymbol{0}^{\top},\
\tau_{1}^{\beta}(\boldsymbol{P}^{\beta}\bomega_1)^{\top}
\Big)^{\top}.
\end{split}
\end{equation*}
and
\[
\boldsymbol{S}_{1,c}-\mathbb{E}(\boldsymbol{S}_{1,c}\mid\boldsymbol{Y})
=
-\frac{1}{\sigma^2}\sum_{i=1}^n\sum_{j=1}^{m_i}
\Big\{\widetilde{\boldsymbol{b}}(t_{ij})^{\top}\boldsymbol{\Theta}\big(\boldsymbol{\xi}_i-\mathbb{E}_i(\boldsymbol{\xi}_i)\big)\Big\}
\Big(\boldsymbol{b}(t_{ij})^{\top},\ \bZ_i^{\top},\ \int_{s\in\mathcal{S}}\boldsymbol{b}^{\beta}(s)^{\top}\mathrm{d}W_i(s)\Big)^{\top}.
\]

Louis' formula gives the observed information matrix as
\begin{equation}\label{eq:louis-def}
\boldsymbol{I}_1
=
-\mathbb{E}\!\left[
\frac{\partial^2\ell_{1,c}}{\partial \boldsymbol{\theta}\,\partial \boldsymbol{\theta}^{\top}}
\Bigm|\boldsymbol{Y}
\right]
-\mathrm{Var}\!\left(
\frac{\partial\ell_{1,c}}{\partial \boldsymbol{\theta}}
\Bigm|\boldsymbol{Y}
\right).
\end{equation}
The complete-data Hessian with respect to $\boldsymbol{\theta}$ does not depend on $\boldsymbol{\xi}_i$%, because
%\[
%\frac{\partial}{\partial\boldsymbol{\theta}}\Big\{
%y_{ij}-\boldsymbol{b}(t_{ij})^{\top}\boldsymbol{\alpha}-\bZ_i^{\top}\bgamma_1-\int_{s\in\mathcal{S}}\boldsymbol{b}^{\beta}(s)^{\top}\bomega_1\,\mathrm{d}W_i(s)
%-\widetilde{\boldsymbol{b}}(t_{ij})^{\top}\boldsymbol{\Theta}\boldsymbol{\xi}_{i}
%\Big\}
%=
%-\Big(\boldsymbol{b}(t_{ij})^{\top},\ \bZ_i^{\top},\ \int_{s\in\mathcal{S}}\boldsymbol{b}^{\beta}(s)^{\top}\mathrm{d}W_i(s)\Big).
%\]
%Therefore, 
and thus taking conditional expectation leaves the Hessian unchanged. The first term in \eqref{eq:louis-def} becomes
\begin{equation}\label{eq:louis-first}
  \begin{split}
&-\mathbb{E}\!\left[
\frac{\partial^2\ell_{1,c}}{\partial \boldsymbol{\theta}\,\partial \boldsymbol{\theta}^{\top}}
\Bigm|\boldsymbol{Y}
\right]\\
=&
\frac{1}{\sigma^2}\sum_{i=1}^n\sum_{j=1}^{m_i}
\Big(\boldsymbol{b}(t_{ij})^{\top},\ \bZ_i^{\top},\ \int_{s\in\mathcal{S}}\boldsymbol{b}^{\beta}(s)^{\top}\mathrm{d}W_i(s)\Big)^{\top}
\Big(\boldsymbol{b}(t_{ij})^{\top},\ \bZ_i^{\top},\ \int_{s\in\mathcal{S}}\boldsymbol{b}^{\beta}(s)^{\top}\mathrm{d}W_i(s)\Big)\\
&\quad+\text{Blockdiag}\{2\tau_{\alpha}\boldsymbol{P},\,\boldsymbol{0},\,2\tau_{1}^{\beta}\boldsymbol{P}^{\beta}\}.
\end{split}
\end{equation}

Now, because $\boldsymbol{\xi}_i$ is shared between measurements within subject $i$, the conditional variance term in \eqref{eq:louis-def} becomes
\begin{equation}\label{eq:louis-second}
\begin{split}
&\mathrm{Var}\!\left(\boldsymbol{S}_{1,c}\mid\boldsymbol{Y}\right)\\
=&
\mathbb{E}\!\left[
(\boldsymbol{S}_{1,c}-\mathbb{E}(\boldsymbol{S}_{1,c}\mid\boldsymbol{Y}))
(\boldsymbol{S}_{1,c}-\mathbb{E}(\boldsymbol{S}_{1,c}\mid\boldsymbol{Y}))^{\!\top}
\ \Big|\ \boldsymbol{Y}\right]
\\
=&
\frac{1}{\sigma^4}\sum_{i=1}^n\sum_{j=1}^{m_i}\sum_{j'=1}^{m_i}
\Big(\boldsymbol{b}(t_{ij})^{\top},\ \bZ_i^{\top},\ \int_{s\in\mathcal{S}}\boldsymbol{b}^{\beta}(s)^{\top}\mathrm{d}W_i(s)\Big)^{\top}
\Big(\boldsymbol{b}(t_{ij'})^{\top},\ \bZ_i^{\top},\ \int_{s\in\mathcal{S}}\boldsymbol{b}^{\beta}(s)^{\top}\mathrm{d}W_i(s)\Big)\\
&\quad\times
\Big[
\widetilde{\boldsymbol{b}}(t_{ij})^{\top}\boldsymbol{\Theta}
\Big\{\mathbb{E}_i(\boldsymbol{\xi}_i\boldsymbol{\xi}_i^{\top})-\mathbb{E}_i(\boldsymbol{\xi}_i)\mathbb{E}_i(\boldsymbol{\xi}_i)^{\top}\Big\}
\boldsymbol{\Theta}^{\top}\widetilde{\boldsymbol{b}}(t_{ij'})
\Big].
\end{split}
\end{equation}
Combining \eqref{eq:louis-def}--\eqref{eq:louis-second}, we have derived the form of $\boldsymbol{I}_1$.

For the survival model,
let $u_i = \bZ_i^{\top}\boldsymbol{\gamma}_{2} + \Big(\int_{s \in \mathcal{S}}  \boldsymbol{b}^{\beta}(s)\mathrm{d}W_i(s) \Big) ^{\top}\bomega_2 + \boldsymbol{\xi}_{i}^{\top}\boldsymbol{\gamma}_{3}$ and $t_v (v =1,\ldots, V)$ are the distinct observed event times.
By differentiating the expected survival log-likelihood with respect to $\boldsymbol{\gamma}$, the score for subject $i$ is $\boldsymbol{s}_i(\boldsymbol{\gamma}) = \{\boldsymbol{s}_i(\boldsymbol{\gamma}_2), \boldsymbol{s}_i(\boldsymbol{\omega}_2), \boldsymbol{s}_i(\boldsymbol{\gamma}_3)\}^\top$, where
$$
\boldsymbol{s}_i(\boldsymbol{\gamma}_2) = \boldsymbol{Z}_i\left\{\Delta_i - 
\sum_{v=1}^V{h}_0(t_v)\mathbb{E}_i 
\exp(u_i){1}_{\{T_i\geq t_v\}}
\right\},
$$
$$
\boldsymbol{s}_i(\boldsymbol{\omega}_2) = \left\{\int_{s \in \mathcal{S}} \boldsymbol{b}^{\beta}(s)\mathrm{d} W_i(s)\right\} \left\{\Delta_i - 
\sum_{v=1}^V{h}_0(t_v)\mathbb{E}_i 
\exp(u_i){1}_{\{T_i\geq t_v\}}
\right\},
$$
and
$$
\boldsymbol{s}_i(\boldsymbol{\gamma}_3)
= \left(\mathbb{E}_i\boldsymbol{\xi}_i\right)
 \left\{\Delta_i - 
\sum_{v=1}^V{h}_0(t_v)\mathbb{E}_i \boldsymbol{\xi}^{\top}_i
\exp(u_i){1}_{\{T_i\geq t_v\}}
\right\}.
$$
%\begin{equation*}
%\begin{split}
 %   &\boldsymbol{s}_i(\boldsymbol{\gamma}) = \Delta_i \left\{\boldsymbol{Z}^{\top}_i, \left(\int_{s \in \mathcal{S}} \boldsymbol{b}^{\beta}(s)\mathrm{d} W_i(s) \right) ^{\top}, \mathbb{E}_i\boldsymbol{\xi}^{\top}_i \right\}^{\top} - \\ & \sum_{v=1}^V{h}_0(t_v)\mathbb{E}_i\Big\{\big(\boldsymbol{Z}^{\top}_i, (\int_{s \in \mathcal{S}}  \boldsymbol{b}^{\beta}(s)\mathrm{d}W_i(s)) ^{\top}, \boldsymbol{\xi}^{\top}_i \big)^{\top}\exp\big(\bZ_i^{\top}\boldsymbol{\gamma}_{2} + \Big(\int_{s \in \mathcal{S}}  \boldsymbol{b}^{\beta}(s)\mathrm{d}W_i(s) \Big) ^{\top}\bomega_2 + \boldsymbol{\xi}_{i}^{\top}\boldsymbol{\gamma}_{3} \big)\Big\}{1}_{\{T_i\geq t_v\}},
%\end{split}
%\end{equation*}
The score $\boldsymbol{S}_{{\boldsymbol{\gamma}}}$ and the information matrix $\boldsymbol{I}_{{\boldsymbol{\gamma}}}$ are calculated as 
$$\boldsymbol{S}_{{\boldsymbol{\gamma}}} = \sum_{i=1}^n\boldsymbol{s}_i(\boldsymbol{\gamma})-(\boldsymbol{0}_P^{\top}, (2\tau_2^{\beta}\boldsymbol{P}^{\beta}\bomega_2)^{\top}, \boldsymbol{0}_L^{\top} )^{\top}$$ and 
$$\boldsymbol{I}_{{\boldsymbol{\gamma}}} = \sum_{i=1}^n\boldsymbol{s}_i(\boldsymbol{\gamma})\boldsymbol{s}_i^{\top}\boldsymbol{\gamma})- \frac{1}{n}\left\{\sum_{i=1}^n\boldsymbol{s}_i(\boldsymbol{\gamma})\right\}\left\{\sum_{i=1}^n\boldsymbol{s}_i^{\top}(\boldsymbol{\gamma})\right\}+ \text{Blockdiag}(\boldsymbol{0}_{P\times P}, 2\tau_2^{\beta}\boldsymbol{P}^{\beta}, \boldsymbol{0}_{L\times L}).$$

%The information matrxi without the 
%$=\sum_{i=1}^n\boldsymbol{s}_i(\boldsymbol{\gamma})\boldsymbol{s}_i^{\top}(\boldsymbol{\gamma})- (\sum_{i=1}^n\boldsymbol{s}_i(\boldsymbol{\gamma}))( \sum_{i=1}^n\boldsymbol{s}_i^{\top}(\boldsymbol{\gamma}))/n$
\section*{Web Appendix E: Additional Data Application Results}\label{app:add.application.result}

\begin{table}[h]
\centering
\caption{Baseline scalar covariates and censoring rates in OPACH data. Race includes white, black and hispanic (w/b/h, \%). Censoring rates are calculated as the proportion of censored observations among the total.}
\begin{tabular}{cc}
        \hline  
         Covariates &  \text{Median (range) / \%}  \\
	\hline 
         \text{Race (w/b/h, \%)}  &  49.9/33.1/17.0 \\
         \text{Age}  &  81.0 (64.0 - 98.0)  \\
         \text{Awake time (hours per day)}   & 15.0 (10.4 - 20.6) \\\hline
         censoring rate & 76.56\% \\
	\hline
\end{tabular}
\label{table:app:baseline.covariates}
\end{table}

Web Table~\ref{table:app:baseline.covariates} provides baseline scalar covariates and censoring rate in OPACH data.

Web Figure~\ref{figure.app.real.baseline.hazard} shows the estimated baseline hazard function in real data application.

\begin{figure}[h]
    \centering
    \includegraphics[width=0.5\textwidth]{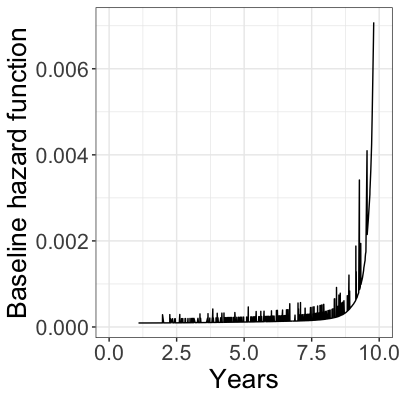}
    \caption{Estimated baseline hazard function ($h_0(t)$) from the FJM applied to real data. }
    \label{figure.app.real.baseline.hazard}
\end{figure}

\section*{Web Appendix F: Additional Simulation Results}\label{app:add.simulation.result}

This section shows additional simulation results to evaluate the model's performance across 100 simulation replicates, with key findings illustrated in several figures to highlight its accuracy and robustness in estimating critical parameters. Web Figure~\ref{fig:simu:eigenvalue} provides boxplots of the estimated eigenvalues. The close agreement between the red lines (true values) and the boxplot medians validates the model's stability and accuracy in capturing variance structures. Web Figure~\ref{fig:simu:gamma} displays the estimated model parameters, including the coefficient vector $\bgamma_1$ in longitudinal model, the Cox regression coefficients $\bgamma_2$ and $\bgamma_3$, and the white noise variance $\sigma^2$. The close alignment of the estimates with true values underscores the model's ability to link longitudinal trajectories to survival outcomes.

These findings collectively affirm the proposed model's effectiveness in handling baseline scalar and functional covariates in longitudinal and survival data, ensuring accurate parameter estimation while maintaining interpretability.

\begin{figure}[h]
    \centering
    \includegraphics[width=0.6\textwidth]{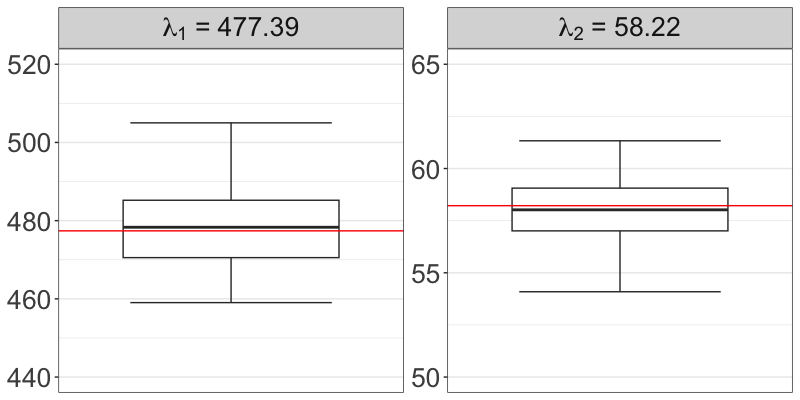}
    \caption{Boxplots of estimated eigenvalues across 100 replicates.}
    \label{fig:simu:eigenvalue}
\end{figure}

\begin{figure}[h]
    \centering
    \includegraphics[width=1\textwidth]{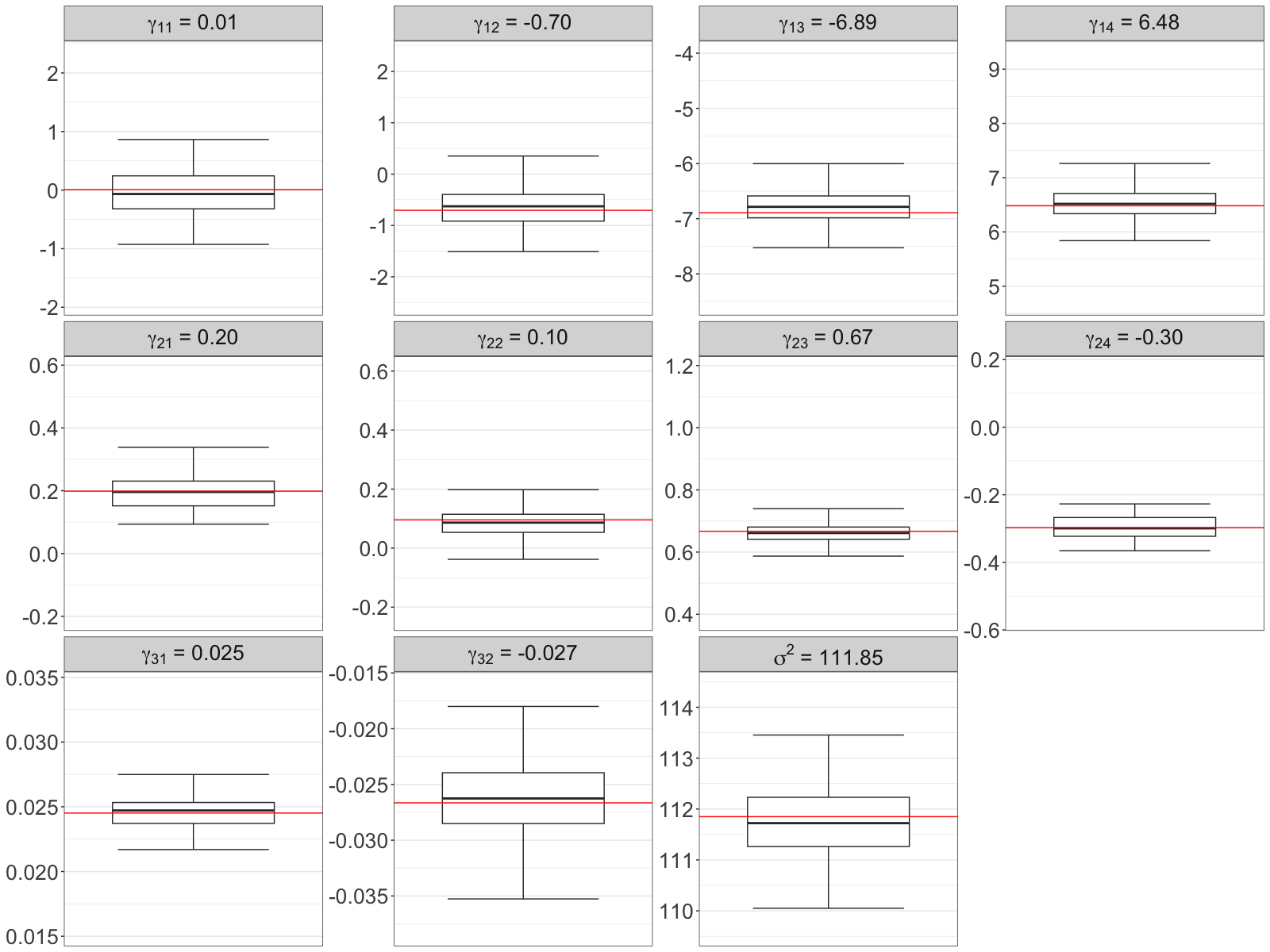}
    \caption{Boxplots of estimated coefficents ($\boldsymbol{\gamma}_1$) for baseline scalar covariates and estimated white noise variance ($\sigma^2$) in longitudinal model and boxplots of Cox regression coefficients ($\boldsymbol{\gamma}_2$, $\boldsymbol{\gamma}_3$) across 100 replicates.}
    \label{fig:simu:gamma}
\end{figure}

\label{lastpage}

\end{document}